\documentclass[aps,prd,showpacs,eqsecnum,amsmath,amssymb,nofootinbib,onecolumn]{revtex4}
\usepackage{graphicx}
\usepackage{subfigure}
\usepackage{array}
\usepackage{multirow}
\usepackage{amsmath}
\usepackage{amssymb}

\usepackage{hyperref}
\usepackage{cleveref}

\begin{document}

\title{\bf Angular and frequency response of the gravitational wave interferometers
in the metric theories of gravity.}

\author{Arkadiusz B{\l}aut}
\affiliation{Institute of Theoretical Physics, University of Wroc\l aw, Wroc\l aw, Poland}

\begin{abstract}
We analyze detector responses of gravitational wave detectors for gravitational waves
with arbitrary polarizations predicted in the metric theories of gravity. We present the
general formulas for the frequency responses valid in various interferometric arrangements
including Michelson, Delay-Line and Fabry-Perot detectors. We analyze the angular
and frequency behavior and the sensitivity patterns of the responses for each
polarization mode.

\end{abstract}


\maketitle

\section{Introduction}
The expected direct detection of the gravitational waves will give the unique opportunity
to study relativistic astrophysical phenomena predicted by the general theory of relativity
and to improve our understanding of the Universe \cite{Thorne95}--\cite{Astro10}. The multiple
detectors are currently operating on Earth (LIGO \cite{LIGO}, VIRGO \cite{VIRGO},
TAMA 300 \cite{TAMA}, GEO 600 \cite{GEO}) and the development of next generation interferometers
is under way. Gravitational waveforms inferred in the process of a data analysis will carry
a remarkably rich information about their sources. For Earth-based detectors they include stellar-mass black-hole binaries, neutron-star-binary mergers, neutron-star normal mode oscillations or core-collapse supernovae. Planned space born detectors, LISA \cite{Bender98}, BBO\cite{Phinney03} and  DECIGO\cite{Kawamura08}, working in the lower frequency bands, below $0.1$ and $10$Hz,
will complement the high-frequency observations on Earth ($10$--$1000$Hz) with signals
from compact binary systems, supermassive black holes mergers, captures of compact stars
by massive black holes and stochastic sources of the Galactic and cosmological origins,
(see \cite{Kawamura11}, \cite{LISArev} and references therein). Despite the possibility
of extending our knowledge of astrophysical phenomena confronting them with models based
on general relativity gravitational waves will be used to test the theory itself potentially
discriminating among various competing alternatives. The present paper aims at this direction by
presenting an analysis of the frequency responses, in low and high frequency domains, of gravitational
wave detectors excited by gravitational waves predicted in a class of metric theories of gravity.

In recent years these theories have attracted much attention as they naturally emerge as the
effective levels of more fundamental models of quantum gravity, string-inspired gravity or in
the dimensional reduction of higher dimensional theories \cite{Buchbinder}. On the other hand
a number of them constitute phenomenological models with an ability to explain some fundamental
astrophysical and cosmological puzzles without evoking dark matter or dark energy \cite{TeVeS}.

In the metric theories of gravity matter fields are minimally and universally coupled
to the physical metric $g$ so the Einstein Equivalence principle is satisfied.
However the dynamics of the physical metric may differ from the one determined by the
standard Einstein-Hilbert action. Metric theories allow for the additional degrees of freedom
in the gravitational sector. They can either be postulated as extra scalar, vector and tensor fields
as in Tensor-Vector-Scalar theories (TeVeS) \cite{TeVeS} or can effectively appear as in the
higher dimensional models \cite{Alesci05} or in four-dimensional theories with modified Hilbert
action like $f(R)$-gravity \cite{Kleinert02}, \cite{Cap02}.

In the metric theories test bodies (e.g. mirrors, beam splitters of detectors) move
along geodesics of the physical metric. From the point of view of the gravitational
waves experiments one is interested in "signatures" that the modified dynamics leaves on the waves.
First, it influences the dynamical processes of wave generations by astrophysical sources.
Thus, for example, it was pointed out in \cite{Sotani09} that oscillation spectra of the neutron star
models in TeVeS theories differ from those expected in general relativity. In \cite{Yunes11} correction
to gravitational waves from nonspinnig black-holes for a large class of alternative theories which
do not admit the Kerr solution were found. The spacetime metric in this case can be studied
with the gravitational waves generated during the inspiral of a compact objects into the massive
black-hole. In \cite{Shibata94} the magnitude of the scalar gravitational waves modeled in the
collapse of a spherical dust in the Brans-Dicke theory was estimated to be large enough to be detected
by the advanced Earth-based interferometers for a stellar-mass supernova in our Galaxy
provided the coupling parameter $\omega_{BD}$ is less than several thousands. The inspiral of a neutron star
into an intermediate-mass black hole in the scala-tensor theories was investigated in \cite{Will05} and the estimated bounds on the parameter $\omega_{BD}$ were given in the context of the proposed LISA mission.

Secondly, in spite of the indirect tests of the strong-field dynamics, one can directly
determine the polarization components of the detected gravitational waves. As predicted in
\cite{Eardley73} a plane gravitational wave in any metric theory can have at most
six polarization modes: scalar longitudinal and scalar transversal modes,
two independent longitudinal-transversal (vector) modes and two transversal tensorial modes.
The linearized theory of TeVeS was investigated in \cite{Sagi10} where the polarization states and
propagation speeds of gravitational waves were found. Similar analysis was performed in
\cite{Jac07}, \cite{Jac04} for theories with a dynamical preferred frame and in \cite{Corda08},
\cite{Cap08} for $f(R)$ theories.
The angular pattern functions for all polarization modes for the Earth-based interferometers were
studied in \cite{Tobar99}. Their detectability with pulsar timing at frequencies $\sim 10^{-8}$Hz
was investigated in \cite{Lee08}. In \cite{Nishizawa09}, \cite{Nishizawa10} a methodology for
the detection and separation of different polarization modes potentially present in the stochastic gravitational-wave background was carried out for Earth-based and space-based detectors.
In \cite{TA10} Tinto et.al. investigated the sensitivity of the LISA detector for all polarization
modes and various interferometric observables in the low and high frequency domains.

Thirdly, in some alternative theories gravitational wave speed may differ from the speed of light.
It can happen when gravity is coupled to a distinguished frame which can be postulated as
a nondynamical background as e.g. in Rosen bimetric theory \cite{Rosen73} or be defined by a dynamical
tensor field as e.g. in "Einstein-aether" theory \cite{Jac07} where the physically
consistent velocities of different polarization modes span two-dimensional manifold.
Another  possibility arises in theories with nonzero mass of gravitons in their
spectra \cite{Goldhaber}. These theories are particularly interesting because mass-like effects
with the associated Compton wavelength of the order of the radius of the visible Universe
might shed a light on the dynamics of the late time cosmic acceleration. Although light gravitons
may arise in several alternative theories (higher dimensional models, $f(R)$-gravities,
bimetric theories) it was already pointed out in \cite{Veltman},\cite{Zakharov} that construction
of a well-defined massive gravity that is consistent with cosmological observations is a nontrivial
open problem. Irrespectively of the underlying mass-generation mechanism it is important that
the gravitational waves exhibit dispersion relation and thus their speed depends on their
energy or wavelength. For the dispersion relation $\omega^2-c^2\,k^2=(m_g c^2/\hbar)^2$, where
$m_g$ is the graviton mass, $\omega$ is the angular frequency of the wave and $k$ its wave
vector the phase velocity is given by
\begin{equation}
\label{eq:KGdr}
v(\omega)=c\left[1-\left(\frac{m_gc^2}{\hbar\,\omega}\right)^2\right]^{-1/2}.
\end{equation}
Assuming the modification of the Newtonian gravitational potential by the Yukawa term,
$e^{-r/\lambda_g}$, where $\lambda_g=h/(m_gc)$ is the graviton Compton wavelength,
one obtains the strongest current bounds for the graviton mass $4.4\times 10^{-22}[eV/c^2]$ from
the solar-system experiments \cite{Talmadge88}, \cite{Will98} and $2.0\times 10^{-29}[eV/c^2]$ from
the galaxy and clusters observations \cite{Goldhaber74}. In turn in a dynamical and
relativistic regime waves propagated by a massive gravitons alter the orbital decay rate of binary
stars. Observations of binary pulsars set the best current 'dynamical' bound to
$7.6\times10^{-20}[eV/c^2]$ \cite{FinnSutton02}.

In the present paper we derive the explicit, general expressions for frequency responses
in various interferometric configurations valid in low and high frequency bands.
Since the status of the future missions and the detector designs are still under scrutiny we derive
the frequency responses for each polarization mode for the Michelson, Delay Line and Fabry-Perot
interferometers. We refer the reader also to \cite{Rakhmanov08} and \cite{Nishizawa08}
were the high-frequency behavior and various interferometric designs were investigated
for the tensor modes. We assume that the interferometers (emitters, beam splitters and mirrors)
move freely along their geodesics. For low frequencies the responses reduce to the known angular pattern functions. We show that our general result applied to the scalar transversal mode agrees with the frequency response derived with a different method in \cite{Nakao01}. To illustrate the results we apply
the obtained explicit formulas in the numerical analysis of the antenna pattern functions for
the stochastic gravitational-wave signal.

The paper is organized as follows. In Sec. \ref{ch:Doppler} we introduce the one-way Doppler shifts
which play the role of basic constituents for the full responses; in Secs. \ref{ch:polar}
and \ref{ch:fresp:pattern} we recall the forms of the gravitational waves in different
polarization modes and give the definitions of the frequency response and antenna pattern function. Finally,
in Sec. \ref{ch:responses} we present the frequency responses in various interferometric
configurations and we analyze their angular and frequency behavior.

For the rest of the paper we take the unit $c=1$.

\section{Doppler tracking}

\label{ch:Doppler}
We consider the Doppler tracking system which consists of two spacecraft,
emitter and receiver of the laser beam, moving freely in a backround geometry with a metric
$$
g_{\mu\nu}=\eta_{\mu\nu} + h_{\mu\nu}
$$
where a small perturbation $h_{\mu\nu}$ represents a gravitational wave passing
through the Minkowski spacetime. Without loss of generality we chose the coordinates
so that the gauge condition $h_{\mu0}=0$ is satisfied.
We also assume that spacecraft are placed at fixed positions of the coordinate system.

The Doppler shift is defined as the frequency fluctuations,
$y_{a,b}(t)\equiv\frac{\nu_b(t,{\bf x}_b)-\nu_a(t_a,{\bf x}_a)}{\nu_a(t_a,{\bf x}_a)}$,
where $\nu_a(t_a,{\bf x}_a)$ is the frequency of the photon emitted from the point ${\bf x}_a$
at the time $t_a$ and $\nu_b(t,{\bf x}_b)$ is the frequency of that photon received
in the point ${\bf x}_b$ at the time $t$. The frequency shift results from
the fluctuations of the phase of the light at the detector which in turn arise due to
time variation of the {\it time-of-flight} of photons moving along the trajectory of the perturbed
geometry. Let us denote $\Delta T_{a,b}(t)$ the time defined by the requirement that the light reaching
${\bf x}_b$ at the coordinate time $t$ left ${\bf x}_a$ at the coordinate time
$t-\Delta T_{a,b}(t)$ so that the phase of the light at
${\bf x}_b$ is $2\pi\nu_{a}(t-\Delta T_{a,b}(t))$ and the Doppler shift is given by $y_{a,b}(t)=-d/dt\,[\Delta T_{a,b}(t)]$.

\begin{figure}[htp]
\begin{center}
\includegraphics[width=20pc]{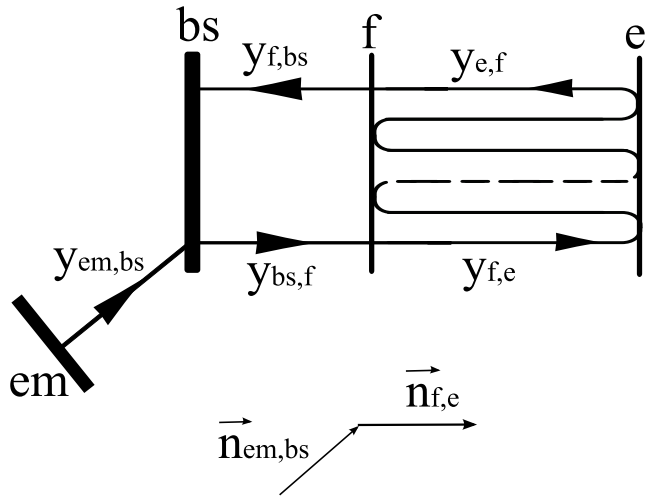}
\end{center}
\caption{The full trajectory of the photon relevant in definition of the response (\ref{eq:dlink}).
The following abbreviations are used: 'em' - emitter, 'bm' - beam splitter, 'f' - front mirror,
'e' - end mirror.}
\label{f:dlink}
\end{figure}

In the Doppler tracking experiment photons follow just the geodesic line between spacecraft however
in other detectors such as Delay-Line (DL) link, Fabry-Perot (FP) cavity or in multi spacecraft
constellations, path lines of photons are in general  more complex. In the case of Michelson (M)
or Delay Line Michelson (DLM) interferometers the emitted laser ray reaches the beam splitter where
it is divided  into two rays which enter into the two arms of the detector and then after taking
single or multiple round trips between mirrors returns to the beam splitter. The typical path of
the photon entering one of the arm in the Delay-Line interferometer is shown in Fig. (\ref{f:dlink}).
The full trajectory consists of the concatenations of several paths:
${\bf x}_{em,bs}$ from the emitter to the beam splitter,
${\bf x}_{bs,f}$ from the beam splitter to the front mirror,
multiple round trip ${\bf x}_{f,e}$, ${\bf x}_{e,f}$
from the front mirror to the end mirror and backward,
and the return pass ${\bf x}_{f,bs}$ from the front mirror to the beam splitter.
Assuming $N$ round trips (with $N$ reflections by the end mirror) one can write the formula
for the cumulative frequency shift of the laser measured at the beam splitter:
\begin{eqnarray}
\label{eq:dlink}
y_{N}(t)&=&
y_{em,bs}(t - 2 L_{f,bs} - 2 N L) + y_{bs,f}(t - L_{f,bs} - 2 N L) +\nonumber\\
&&
\sum_{i=1}^{N}y_{f,e}(t - L_{f,bs} - (2 i -1) L) +
\sum_{i=1}^{N}y_{e,f}(t - L_{f,bs} - 2 (i-1) L) +\nonumber\\
&&
y_{f,bs}(t),
\end{eqnarray}
where we have shortened the notation for the length between the two mirrors, $L_{f,e}=L_{e,f}\equiv L$.
The total frequency fluctuation $y_{N}(t)$ in the formula (\ref{eq:dlink}) consists of the sum of the
terms $y_{em,bs}$, $y_{bs,f}$, $y_{f,e}$, $y_{e,f}$, $y_{f,bs}$ which determine the contributions
of the corresponding Doppler shifts along the paths
${\bf x}_{em,bs}$, ${\bf x}_{bs,f}$, ${\bf x}_{f,e}$, ${\bf x}_{f,e}$, ${\bf x}_{f,bs}$
respectively and are taken at the proper time delays.

\section{Polarization states}

\label{ch:polar}
\begin{figure}[htp]
\begin{center}
\includegraphics[width=16pc]{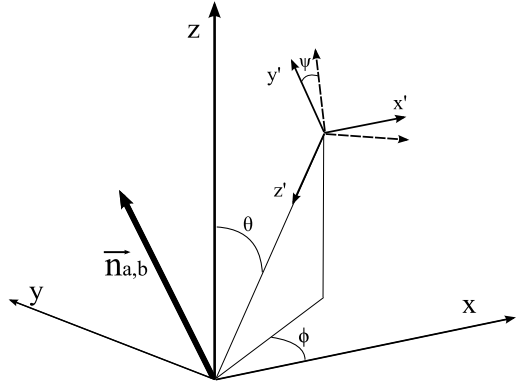}
\end{center}
\caption{Convention used in the definition of the polar angle $\psi$ and the source
frame $\{x',y',z'\}$. In that frame gravitational waves take their canonical forms
discussed in Sec. \ref{ch:polar}. The dashed unit vectors are tangent to the lines of constant
$\theta$ and $\phi$; $z'$ axis lies along the direction of the wave propagation.
It is assumed that the origins of the frames $\{x,y,z\}$ and $\{x',y',z'\}$ coincide.
The unit vector ${\bf n}_{a,b}$ indicates the orientation of the detector's arm.}
\label{f:frame}
\end{figure}

To characterize the polarization states of the perturbation tensor we expand a plane
gravitational wave moving in the direction ${\bf\Omega}$ with respect to the polarization
tensors, ${\rm \bf h}(t,{\bf x})=\sum_{\pi}\,{\rm h}_{\pi}(t,{\bf x}){\boldsymbol \epsilon}^{\pi}$.
We assume that in the source frame $\{x',y',z'\}$ (see Fig.\ref{f:frame}) components
of ${\boldsymbol \epsilon}^{\pi}$ take the following forms,
\begin{eqnarray}
\label{eq:polarizations:s}
\tilde{\epsilon}^{sl} & = &
\left(
\begin{array}{ccc}
0 & 0 & 0 \\
0 & 0 & 0 \\
0 & 0 & 1
\end{array}
\right),\qquad
\tilde{\epsilon}^{st} =
\left(
\begin{array}{ccc}
1 & 0 & 0 \\
0 & 1 & 0 \\
0 & 0 & 0
\end{array}
\right),\\
\label{eq:polarizations:v}
\tilde{\epsilon}^{vx} & = &
\left(
\begin{array}{ccc}
0 & 0 & 1 \\
0 & 0 & 0 \\
1 & 0 & 0
\end{array}
\right),\qquad
\tilde{\epsilon}^{vy} =
\left(
\begin{array}{ccc}
0 & 0 & 0 \\
0 & 0 & 1 \\
0 & 1 & 0
\end{array}
\right),\\
\label{eq:polarizations:t}
\tilde{\epsilon}^{tp} & = &
\left(
\begin{array}{ccc}
1 & 0 & 0 \\
0 & -1 & 0 \\
0 & 0 & 0
\end{array}
\right),\qquad
\tilde{\epsilon}^{tc} =
\left(
\begin{array}{ccc}
0 & 1 & 0 \\
1 & 0 & 0 \\
0 & 0 & 0
\end{array}
\right)
\end{eqnarray}
defining two scalar (longitudinal '$sl$' and transversal '$st$'), two vectorial (longitudinal-$x$
'$vx$' and longitudinal-$y$ '$vy$') and two tensorial (transversal-$+$ '$tp$' and transversal-$\times$
'$tc$') modes respectively.
For each polarization mode $\pi$ we define function $F^{\pi}({\bf n}_{a,b};\theta,\phi,\psi)$
which depends on the unit vector ${\bf n}_{a,b}$ determining the attitude of the arm of the
interferometer and the angles $\theta$, $\phi$, $\psi$ specifying orientation of the source
frame (see Fig. \ref{f:frame}):
\begin{eqnarray}
\label{eq:F}
F^{\pi}({\bf n}_{a,b};\theta,\phi,\psi) & := &
\frac12 {\bf n}_{a,b}\otimes{\bf n}_{a,b}:{\boldsymbol \epsilon}^{\pi}
= \frac12 \tilde{n}_{a,b}^T \;\tilde{\epsilon}^{\pi} \;\tilde{n}_{a,b},
\end{eqnarray}
where the colon denotes the double contraction, $\tilde{n}_{a,b}$ is the column-vector of
the components of ${\bf n}_{a,b}$ in the source frame and $T$ stands for a matrix transposition.
Functions $F^{\pi}$ given in Eqs.(\ref{eq:F}) are called the angular pattern functions
and play the role of the antenna pattern functions (for the one-arm detector) in the long-wavelength
(LW) approximation which is defined as the leading-term in the $\omega L\rightarrow 0$ limit of the
exact response.

In turn we define functions $u$ and $v$ by the decompositions
\begin{align}
\label{eq:uv}
F^{\pi} & =:  u &&\text{for $\pi=sl,\;st$}\\
F^{\pi} & =:  u\cos{\psi} + v\sin{\psi} &&\text{for $\pi=xz,\;yz$}\\
F^{\pi} & =:  u\cos{2\psi} + v\sin{2\psi} &&\text{for $\pi=tp,\;tc$}
\end{align}

reflecting spin-$0$, spin-$1$ and spin-$2$ contents of the corresponding perturbations.

\section{Frequency response and antenna pattern function}
\label{ch:fresp:pattern}
In this section we present explicit expressions for the responses $\Delta T_{a,b}(t)$
and $y_{a,b}(t)$ and recall the definitions of the frequency response and antenna pattern
function for one-arm interferometer.

\subsection{Deterministic signal}

First we assume that the perturbed geometry is defined by a plane wave moving
with the velocity $v$ in the direction determined by the unit vector ${\bf \Omega}$:
$\;\rm{\bf h}(t,{\bf x})=\rm{\bf h}(t-{\bf\Omega}\cdot{\bf x}/v)$.
The unit vector ${\bf n}_{a,b}$ is oriented from the emitter to the receiver
and ${\rm L}_{a,b}$ is the separation between the spacecraft.
Following Finn \cite{Finn09} we express $\Delta T_{a,b}(t)$ as
\begin{eqnarray}
\label{eq:DTab}
\Delta T_{a,b}(t) & = & L_{a,b}+\frac{{\bf n}_{a,b}\otimes {\bf n}_{a,b}}{2}:\int_{t-L}^{t}
{\rm\bf h}[t_0(\lambda),{\bf x}_0(\lambda)]\,d\lambda\\
& = &
L_{a,b} + \frac{{\bf n}_{a,b}\otimes {\bf n}_{a,b}}{2(1-\frac{{\bf\Omega}\cdot{\bf n}_{a,b}}{v})}:\int_{u(t-L_{a,b})}^{u(t)}
{\rm\bf h}(u)\,du,\nonumber
\end{eqnarray}
where $u(\lambda)=t_0(\lambda)-{\bf\Omega}\cdot{\bf x}_0(\lambda)/v$
and the integrals on the right hand sides are taken along the {\it unperturbed} trajectory
\begin{eqnarray}
\label{eq:b2}
t_0(\lambda) & = & \lambda,\\
{\bf x}_0(\lambda)& = & {\bf x}_{b}-{\bf n}_{a,b}(t-\lambda).
\end{eqnarray}
From this we obtain the Doppler shift
\begin{equation}
\label{eq:yab}
y_{a,b}(t)=-\frac{d}{dt}\Delta T_{a,b}(t)=
\frac{{\bf n}_{a,b}\otimes {\bf n}_{a,b}}{2(1-\frac{{\bf\Omega}\cdot{\bf n}_{a,b}}{v})}:
\left[
{\rm\bf h}(t - L_{a,b} - {\bf\Omega}\cdot{\bf x}_a/v) - {\rm\bf h}(t - {\bf\Omega}\cdot{\bf x}_b/v)
\right].
\end{equation}
The last formula can also be obtained by the algebraic method making use of the Killing vectors
of the perturbed metric (\cite{Wahlquist87},\cite{Hellings78}). It can be used to obtain the responses
in more complex interferometric configurations and we will apply it to get the the full detector response (\ref{eq:dlink}).

More generally one has a superposition of monochromatic plane waves with
different velocities specified by a dispersion relation $v_{\omega}$
so that the retarded time $u_{\omega}=t-{\bf \Omega}\cdot{\bf x}/v_{\omega}$
can be different for each frequency component:
\begin{equation}
\label{b3}
{\rm\bf h}(t,{\bf x})=\int_{-\infty}^{\infty}\tilde{\rm\bf h}(\omega)
e^{i\omega(t-{\bf\Omega}\cdot{\bf x}/v_{\omega})}\,d\omega.
\end{equation}
In this case the time-dependent part of the travel time reads
\begin{eqnarray}
\label{eq:b4}
\Delta T_{a,b}(t) & = &
\int_{-\infty}^{\infty}
{\bf D}(\omega;{\bf n}_{a,b},{\bf\Omega}):\tilde{\rm\bf h}(\omega)\,e^{i \omega u_{\omega}(t)}\,d\omega,
\end{eqnarray}
where $u_{\omega}(t)=t-{\bf \Omega}\cdot{\bf x}_b/v_{\omega}$.
The one-arm detector tensor is given by
\begin{equation}
\label{eq:dten}
{\bf D}(\omega;{\bf n},{\bf\Omega}) = \frac{L}{2}{\bf n}\otimes{\bf n}\,
{\cal T}(\omega;{\bf x},{\bf \Omega}),\qquad\qquad {\bf x}=L{\bf n}
\end{equation}
and the one-way transfer function reads
\begin{equation}
\label{eq:T}
{\cal T}(\omega;{\bf x},{\bf \Omega})=
\text{sinc}{[\frac{\omega L}{2}(1 - {\bf \Omega}\cdot{\bf n}/v_{\omega})]}\,
e^{-\frac{i \,\omega L}{2}(1 - {\bf \Omega}\cdot{\bf n}/v_{\omega})}.
\end{equation}

In next chapters we will consider the angular and frequency characteristics of interferometers
for plane waves in a definite polarization states,
$\;\tilde{\bf h}(\omega)=\tilde{\rm h}_{\pi}(\omega)\,{\boldsymbol \epsilon}^{\pi}$.
We define the frequency response function $H(\omega)$ by
\begin{equation}
\label{eq:fresp}
\Delta T(t) = i\,\int_{-\infty}^{\infty}\tilde{\rm h}_{\pi}(\omega)
H(\omega)e^{i \omega u_{\omega}(t)}\,d\omega.
\end{equation}
Thus in the case at hand of one-arm interferometer
$H(\omega)=-i\,L_{a,b}\,{\cal T}(\omega;{\bf x}_{a,b},{\bf \Omega})F^{\pi}({\bf n}_{a,b})$.
If one takes the Doppler shift as a basic observable the frequency response would be
given by $-i\omega H(\omega)$. To compare with the frequency responses given
in a literature in the paper we state our results in terms of  $H(\omega)$.

The above formulas, Eqs. (\ref{eq:DTab}),(\ref{eq:yab}),(\ref{eq:dten}),({\ref{eq:T}}),
are valid for an arbitrary frequency-dependent velocity $v_{\omega}$. They show that effects of
$v_{\omega}\neq 1$ can generally be taken into account by replacing $\Omega$
with $\Omega/v_{\omega}$ in the detector responses. This shows that the corrections to the
detector response are due to the offset in the time spent by the perturbation in the
detector's arm. These corrections are negligible in theories of massive graviton if one assumes
the current upper limits on the graviton mass. (But in this case one should exclude in the detector
response frequencies below $(m_gc^2)/h$ which is out of the observable band of any present or
planned detector.) For these two reasons in the rest of the paper we put $v_{\omega}=1$.
We note however that even if different wavelengths of a gravitational wave travel with slightly
different velocities they can give in the case of distant sources detectable corrections
in the evolution of the signal's phase \cite{Will98}.

\subsection{Stochastic signal}

We will also consider a stochastic signal of the background field formed by a superposition
of the plane monochromatic waves
\begin{equation}
\label{eq:s1}
{\bf h}(t,{\bf x})=\sum_{\pi}\int_{-\infty}^{\infty}\,
d\omega
\int_{S^2}\,
d{\bf\Omega}\,
\tilde{\rm h}_{\pi}(\omega,{\bf\Omega}){\boldsymbol \epsilon}^{\pi}(\bf\Omega)\,e^{i\omega(t-{\bf\Omega}\cdot{\bf x})}
\end{equation}
where the zero-mean random Fourier amplitudes $\tilde{h}_{\pi}$ are characterized by
the power spectral density $S(\omega,{\bf\Omega})$:
\begin{equation}
\label{eq:s2}
\langle \tilde{\rm h}^{*}_{\pi}(\omega,{\bf\Omega})
\tilde{\rm h}_{\pi'}(\omega',{\bf\Omega}') \rangle =
\frac12 S(\omega,{\bf\Omega}) \delta_{\pi\pi'} \delta(\omega-\omega')
\frac{\delta^2({\bf\Omega},{\bf\Omega}')}{4\pi},\qquad\qquad
S(-\omega,{\bf\Omega})=S(\omega,{\bf\Omega}),
\end{equation}
and star denotes complex conjugation.

In the case of stochastic signals the quantity one wants to detect is the spectral density
$S(\omega,{\bf\Omega})$ of the possibly anisotropic background.
Since the mean value of the signal measured by the interferometer vanishes one needs
to perform the correlation analysis so the observable quantity are correlations
$\langle s_I(t)s_J(t) \rangle$ of the two signals, $s_I$ and $s_J$, measured by detectors
located at points ${\bf x}_I$ and ${\bf x}_J$. Depending on the search strategy self-correlations
($s_I=s_J$) or cross-correlations can be used. For the time-of-flight $\Delta T(t)$ measurements
using Eqs. (\ref{eq:b4}),(\ref{eq:s1}),(\ref{eq:s2}) one obtains
\begin{equation}
\label{eq:s3}
\langle s_I(t)s_J(t) \rangle =
\frac12\int_{-\infty}^{\infty}\int_{S^2}
S(\omega,{\Omega}){\cal F}_{IJ}(\omega,{\bf\Omega})\frac{d\bf\Omega}{4\pi}d\omega,
\end{equation}
so that the response of the detector can be characterized by the antenna pattern function
\begin{equation}
{\cal F}_{IJ}(\omega,{\bf\Omega})=
e^{i\omega\, {\bf\Omega}\cdot({\bf x}_J-{\bf x}_I)}\sum_{\pi}
\left[{\bf D}_I(\omega,{\bf\Omega}):{\boldsymbol \epsilon}^{\pi}(\bf\Omega)\right]
\left[{\bf D}_J^*(\omega,{\bf\Omega}):{\boldsymbol \epsilon}^{\pi}(\bf\Omega)\right],
\end{equation}
where ${\bf D}_I$ and ${\bf D}_J$ are the detector tensors. Explicit expressions
for ${\bf D}:{\boldsymbol \epsilon}^{\pi}$ will be given in Sec.\ref{ch:triangular}.

\section{Detector responses}
\label{ch:responses}

\subsection{Single arm}
\label{ch:singlearm}
For the derivation of the $N$-round-trip response $y_{N}(t)$ we refer the
reader to Appendix \ref{app:DL}. The frequency response function for the single-arm
detector working in the Delay Line setup as in Fig.\ref{f:dlink} with $N$ round trips
has the following explicit form
\begin{eqnarray}
\label{eq:dlink:DL}
H_{N}(\omega) & = &
-i\left\{L_{bs,f}\,{\cal T}(\omega;{\bf x}_{f,bs},{\bf \Omega})  +
\right.\\
&&
2\,N\,L\,{\cal T}_N(\omega;{\bf x},{\bf \Omega})\,e^{-i \omega L_{bs,f}(1+c)}+\nonumber\\
&&
\left.\left.
L_{bs,f}\,{\cal T}(\omega;{\bf x}_{bs,f},{\bf \Omega}) e^{-i \omega L_{bs,f}(1+c)}
e^{-2 N i \;\omega L}\right]
\right\}F^{\pi}({\bf n})\nonumber\\
&&
-i
L_{em,bs}\,{\cal T}(\omega;{\bf x}_{em,bs},{\bf \Omega}) e^{-2i \omega L_{bs,f}}e^{-2i N \omega L}
F^{\pi}({\bf n}_{em,bs})\nonumber,
\end{eqnarray}
where ${\bf x}\equiv{\bf x}_{f,e}$, ${\bf n}\equiv{\bf n}_{f,e}$,
$c\equiv{\bf \Omega}\cdot{\bf n}$ and we omitted the angular variables in $F^{\pi}$.
The transfer functions
${\cal T}_N$ for a multiple round trips entering Eq.(\ref{eq:dlink:DL}) is defined
in Appendix \ref{app:DL}.

For a Fabry-Perot cavity we assume that the reflection coefficients
of the front mirror inside and outside cavity are $\rho$ and $-\rho$ respectively,
the transmission coefficient of the front mirror is given by $\nu$;
we also assume that the end mirror has a perfect reflectivity and the system is loss-free,
i.e. $\rho^2+\nu^2=1$. In this case the signal at the beam splitter is given by
\begin{eqnarray}
\label{eq:dlink:FPy}
y_{FP}(t) & = &
-\rho\,y_{(0)}(t) + \nu^2\sum_{N=1}^{\infty}\rho^{N-1} y_{(N)}(t)
\end{eqnarray}
and the frequency response
$H_{FP}(\omega)$ reads
\begin{eqnarray}
\label{eq:dlink:FP}
H_{FP}(\omega) & = &
-i\left\{
L_{bs,f}\,{\cal T}(\omega;{\bf x}_{f,bs},{\bf \Omega}) +
\right.\\
&&
2\,L\,\frac{1+\rho}{1-\rho}\,{\cal T}_{FP}(\omega;{\bf x},{\bf \Omega})\,
e^{-i \omega L_{bs,f}(1+c)} +\nonumber\\
&&
\left.\left.
L_{bs,f}\,{\cal T}(\omega;{\bf x}_{bs,f},{\bf \Omega}) e^{-i \omega L_{bs,f}(1+c)}
\frac{e^{-2i \omega L}-\rho}{1-\rho e^{-2i \omega L}}
\right]
\right\}F^{\pi}({\bf n})\nonumber\\
&&
- i\,
L_{em,bs}\,{\cal T}({\bf x}_{em,bs}) e^{-2i \omega L_{bs,f}}
\frac{e^{-2i \omega L}-\rho}{1-\rho e^{-2i \omega L}}
F^{\pi}({\bf n}_{em,bs})\nonumber,
\end{eqnarray}
where the transfer function ${\cal T}_{FP}$ for a Fabry-Perot cavity is defined in Appendix \ref{app:DL}.

In the approximation when $\omega L_{bs,f}\ll 1$, $\omega L_{em,bs}\ll 1$,
$L_{bs,f}\ll L$, $L_{em,bs}\ll L$ the frequency responses (\ref{eq:dlink:DL})
and (\ref{eq:dlink:FP}) read
\begin{align}
\label{eq:dlink:apprN}
H_{N}(\omega) & =  H_{N}^{LW}(\omega)\;{\cal T}_{N}
,
\\
\label{eq:dlink:apprFP}
H_{FP}(\omega) & =  H_{1}^{LW}(\omega)\;\frac{1+\rho}{1-\rho}{\cal T}_{FP}
,
\end{align}
where the frequency response in the LW limit is given by
\begin{equation}
\label{eq:dlink:lw}
H_{N}^{LW}(\omega) = -2\,i\,N\,L\,F^{\pi}({\bf n})
\qquad\qquad (L_{em,bs}\ll L, \; L_{bs,f}\ll L, \; \omega L\ll 1).
\end{equation}
We see from Eqs.(\ref{eq:dlink:apprN})-(\ref{eq:dlink:lw}) and
(\ref{eq:single})-(\ref{eq:TFP}) that the frequency responses for the single round trip,
DL and FP arms are related by the same simple multiplication factors independently on the type of
polarization mode. Likewise the response function for a single round trip defines the
upper envelope (modulo the amplification factor $(1+\rho)/(1-\rho)$) for the response of FP
cavity and the two are equal at the resonant frequencies $\omega L=n\pi$  which was observed in
\cite{Schilling97} in the case of tensorial polarizations.

\begin{figure}[htp]
\begin{center}
\subfigure{
\includegraphics[width=12pc]{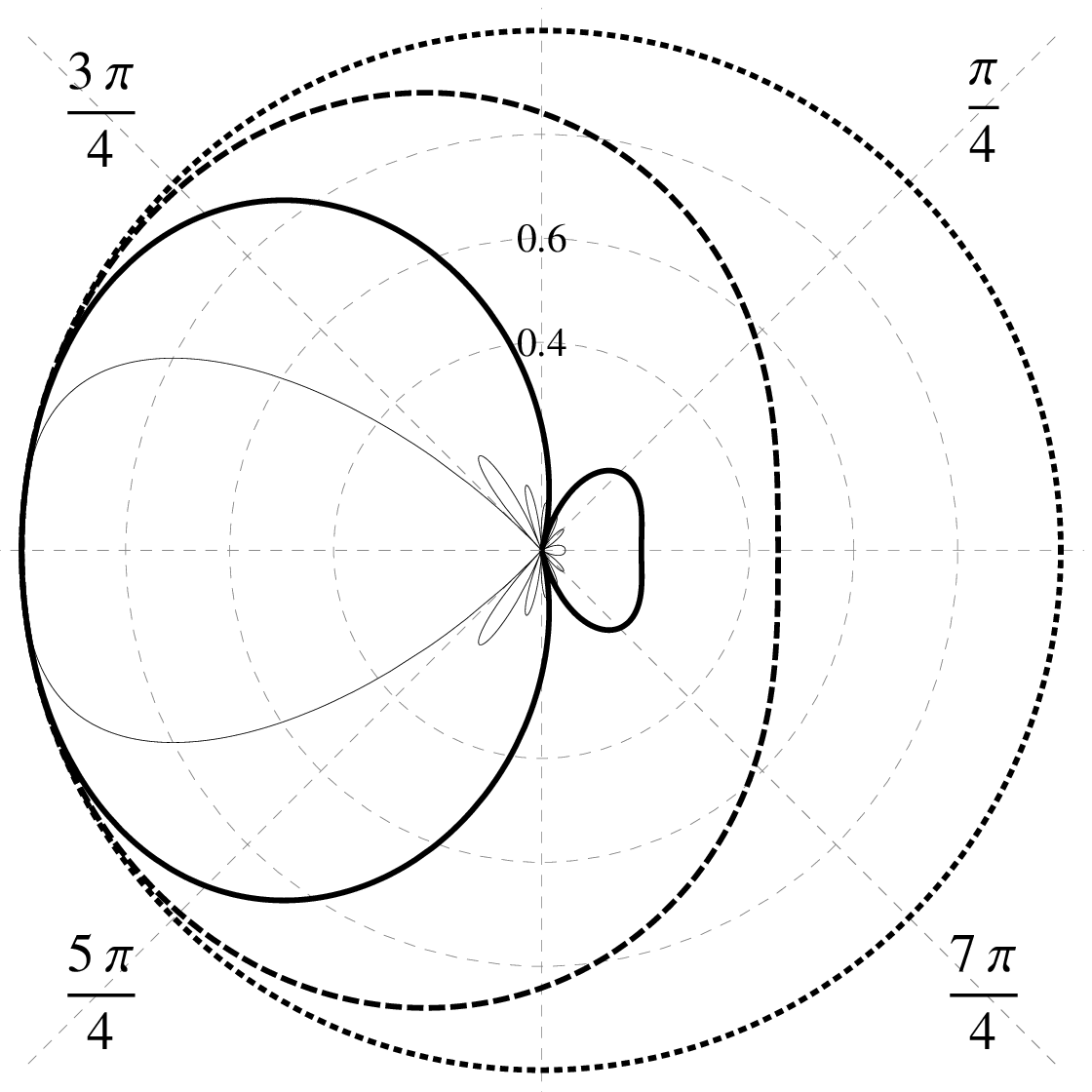}}
\subfigure{
\includegraphics[width=12pc]{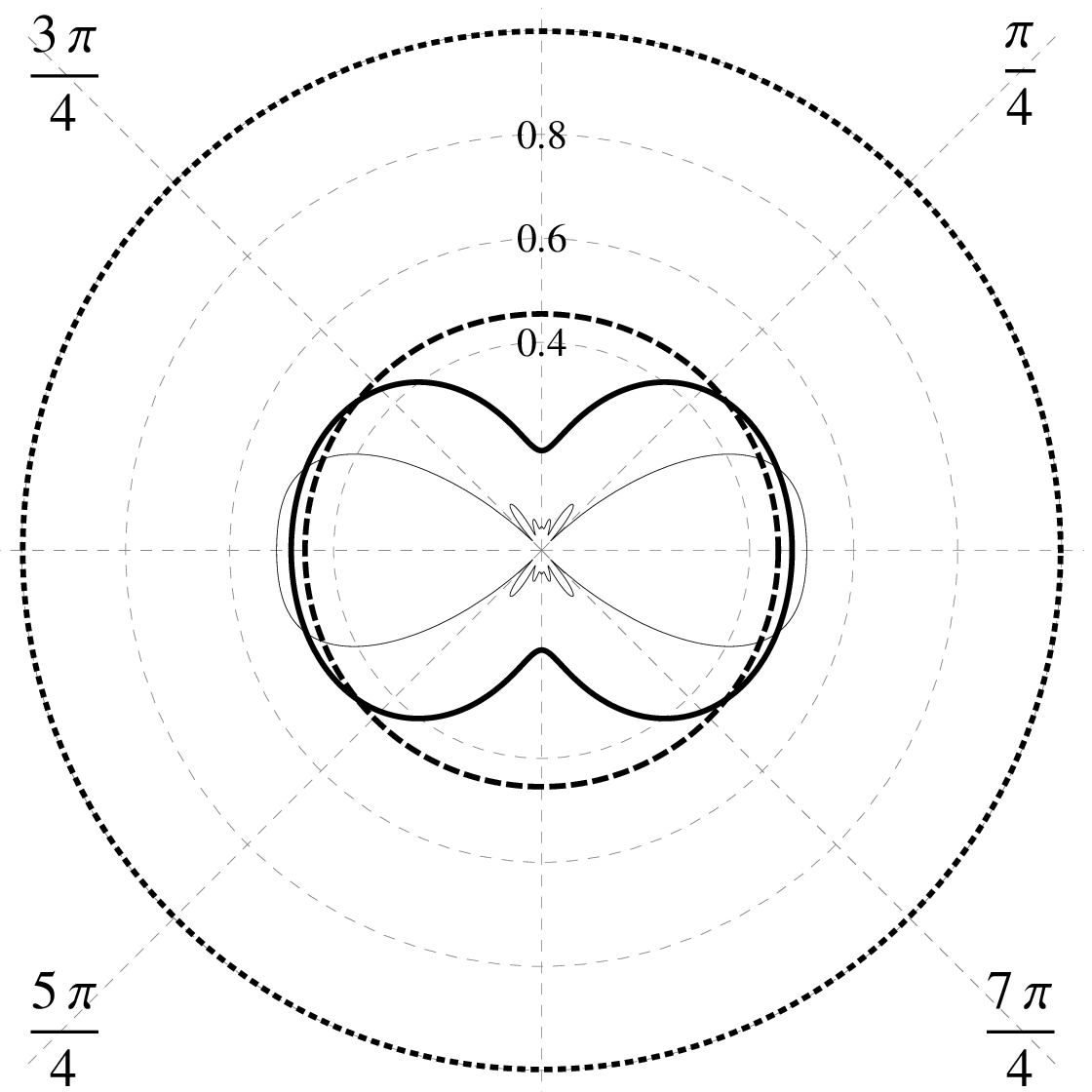}}
\subfigure{
\includegraphics[width=18pc]{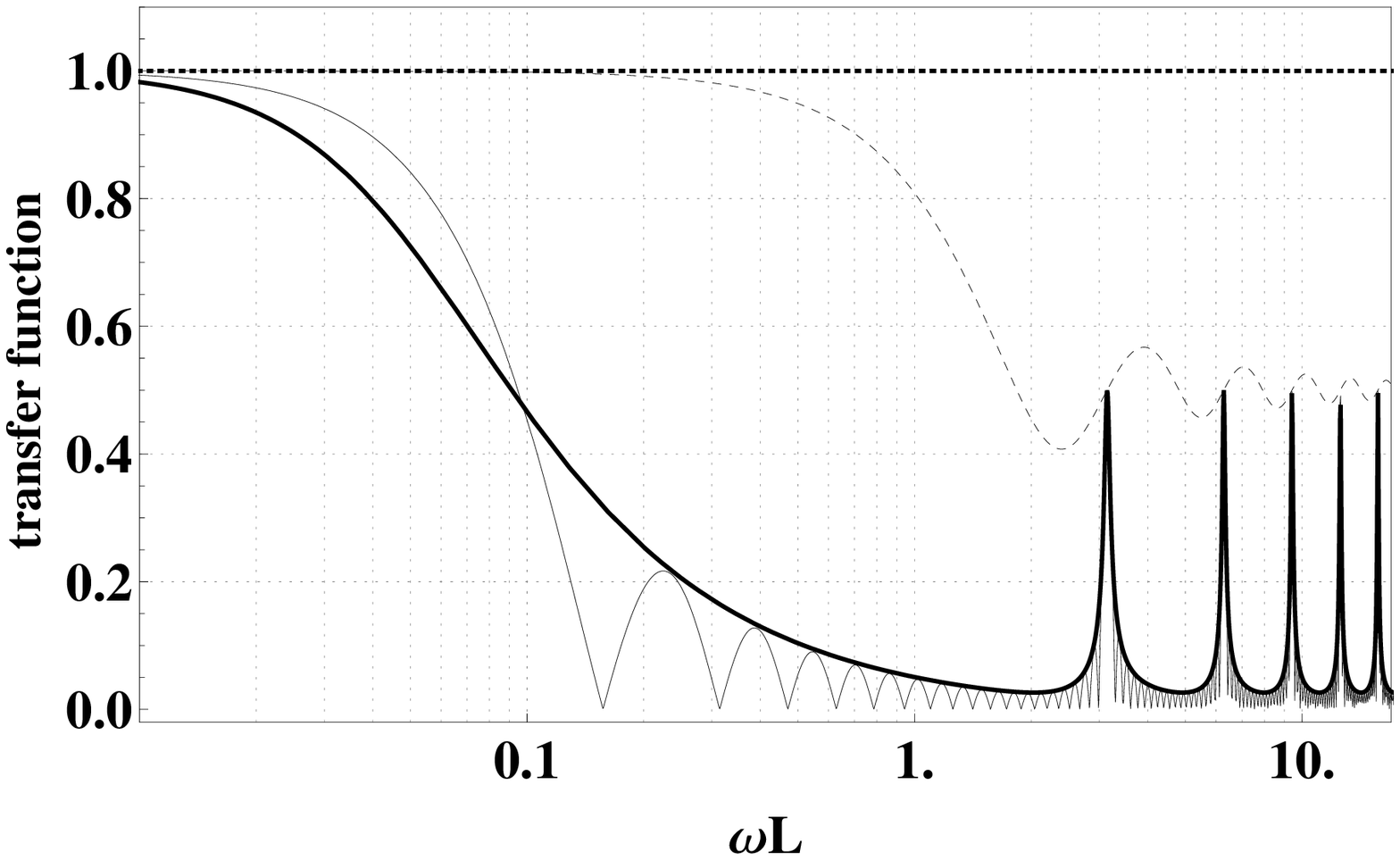}}
\end{center}
\caption{Angular dependence of the magnitude of the transfer functions ${\cal T}$ (upper left)
and ${\cal T}_{1}$ (upper right) for $\omega L_{a,b}$ equal $0.1$ (dotted line), $2$ (dashed line),
$5$ (thick line) and $20$ (thin line). Lower plot shows the dependence on the frequency
of ${\cal T}$ (dotted curve), ${\cal T}_{1}$ (dashed curve), ${\cal T}_{20}$ (thin curve)
and ${\cal T}_{FP}$ for $\rho=0.9$ (thick curve) at the point $\vartheta=-\pi$.
\label{f:Tf}}
\end{figure}

The high-frequency behavior of the single arm interferometers is determined
by the transfer functions  ${\cal T}$, ${\cal T}_{N}$, ${\cal T}_{FP}$ given in Eqs.(\ref{eq:T}),
(\ref{eq:TDL}), (\ref{eq:TFP}), which depend on the frequency and the angle $\vartheta$
between the link and the direction of the wave propagation ($\vartheta$ should not be confused
with $\theta$ defined in Fig.\ref{f:frame}). All three functions are normalized
so in the LW limit they tend to unity. At high-frequency they define functions that vanish
as $1/(\omega L)$ for all $\vartheta$'s except $\vartheta=\pi$ where ${\cal T}(\vartheta)=1$ and
$\vartheta=0$, $\pi$ where ${\cal T}_{N}(\vartheta)\rightarrow1/2$ and ${\cal T}_{FP}(\vartheta)$
oscillates between $1$ and $(1-\rho)/(1+\rho)$. In Fig.\ref{f:Tf} we show $\vartheta$-dependence
of the transfer functions for some selected frequencies and their dependence on the frequency at
$\vartheta=-\pi$ (see similar plots in \cite{Schilling97} for different values of $\vartheta$).
Using now the functions $u$ and $v$ (see Eqs.(\ref{eq:uv:sl})-(\ref{eq:uv:tc}) in Appendix \ref{app:uv})
that determine the shape of the detector response in the low frequency band from Eqs.(\ref{eq:dlink:DL})
and (\ref{eq:dlink:FP}) one can infer the single-arm detector responses for each polarization mode.
We note here the distinctive frequency dependence of the scalar longitudinal mode
at $\vartheta_0=\pm\pi$. This mode alone has nonvanishing antenna pattern function $u_{sl}$
at $\vartheta_0$ and consequently the frequency response for the wave coming from that direction
approaches constant nonzero value at high frequency.
This effect was already noticed in \cite{TA10} in the study of sensitivity of the LISA array.
One can better understand it by considering the change of the time of arrivals of photons moving
from the emitter to the receiver in a background of gravitational wave traveling in the same direction.
In this particular case one can show using the geodesic equation of motion that the trajectory
of photons is unaffected for all polarizations given in (\ref{eq:polarizations:s}-\ref{eq:polarizations:t}) save
the scalar longitudinal mode in which case the delay in the time of arrival of photons
is a frequency independent constant and thus equal to its LW limit.
Heuristically then one can admit the view of photons surfing on the gravitational wave
and perceiving it as a constant gravitational field.

\subsection{Two arms}
\label{ch:frresp}
\begin{figure}[htp]
\begin{center}
\includegraphics[width=30pc]{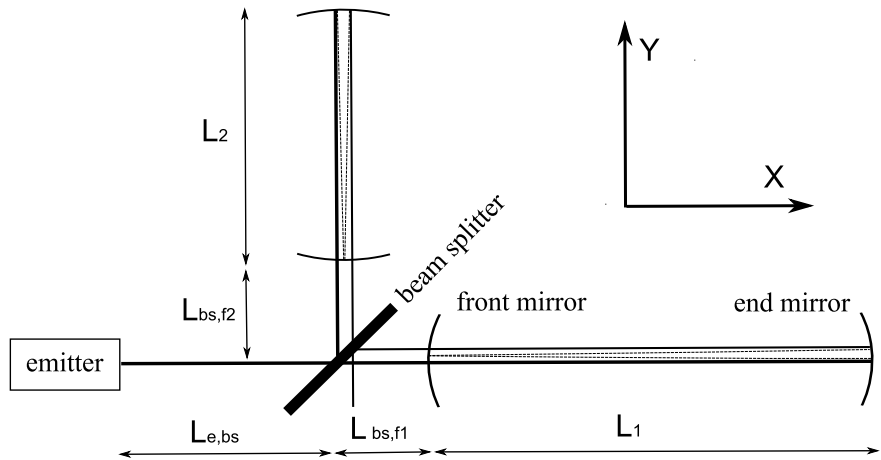}
\end{center}
\caption{The configuration and orientation of the two arms of the Michelson, Delay-Line and Fabry-Perot interferometers discussed in Sec.{\ref{ch:frresp}}.}
\label{f:frame2}
\end{figure}

The response for the differential interferometers can be obtained as the difference of two
responses (\ref{eq:dlink:DL}) or (\ref{eq:dlink:FP}) given in Sec.\ref{ch:singlearm}
defined for single-arm interferometers. Assuming that $\omega L_{bs,f}\ll 1$, $\omega L_{em,bs}\ll 1$, $L_{bs,f}\ll L$, $L_{em,bs}\ll L$
for both arms we obtain for the Michelson interferometer
\begin{eqnarray}
\label{eq:M:appr}
H_{M}(\omega) & = &
H^{(1)}_1(\omega) - H^{(2)}_1(\omega)
\end{eqnarray}
where we have introduced the frequency responses for the two arms:
\begin{eqnarray}
\label{eq:M1:appr}
H^{(i)}_1(\omega) & = &
\frac{- 1 + e^{-2 i \omega L_i} + \left(1 + e^{-2 i \omega L_i} - 2 e^{-i \omega L_i(1 + c_i)}\right)c_i}
{\omega\left(1 - c_i^2\right)}\,F^{\pi}({\bf n}_i)
\qquad (i\,=\,1,\,2\,).
\end{eqnarray}
For the DLM and Fabry-Perot Michelson (FPM) detectors we then have
\begin{eqnarray}
\label{eq:DLM:appr}
H_{DLM}(\omega) & = &
\frac{\sin{N \omega L_1}}{\sin{\omega L_1}}\,e^{-i \omega L_1(N - 1)}\,H^{(1)}_1(\omega)
\quad - \quad  \left( 1\leftrightarrow 2 \right)\\
H_{FPM}(\omega) & = &
\frac{1 + \rho}{1 - \rho\,e^{-2 i \omega L_1}}\,H^{(1)}_1(\omega)
\quad - \quad  \left( 1\leftrightarrow 2 \right).
\end{eqnarray}
When the two arms of the interferometer lie along the $x$ and $y$ axes as in
Fig.\ref{f:frame2} we have $c_1=-\cos{\phi}\sin{\theta}$ and $c_2=-\sin{\phi}\sin{\theta}$.
The general case with arbitrarily oriented arms is set out in Appendix \ref{ch:DLM}.
We note that the exact formulas (\ref{eq:DLM:full}) and (\ref{eq:FPM:full})
given in Appendix \ref{ch:DLM} applied to ${\bf n}_1={\bf e}_x$ and
${\bf n}_1={\bf e}_y$ reproduce, in the case of the scalar transversal mode,
all the results derived in \cite{Nakao01}. In that paper the frequency response was obtained in
rigorous analysis of the geodesic equation of motion of the emitter, beam splitters and
mirrors and the Maxwell field equation for the propagation of the laser ray. Moreover the
calculation were carried out in a different gauge so the compatibility of the two methods
comprises a relevant consistency check.

\begin{figure}[htp]
\begin{center}
\subfigure{
\includegraphics[width=35pc]{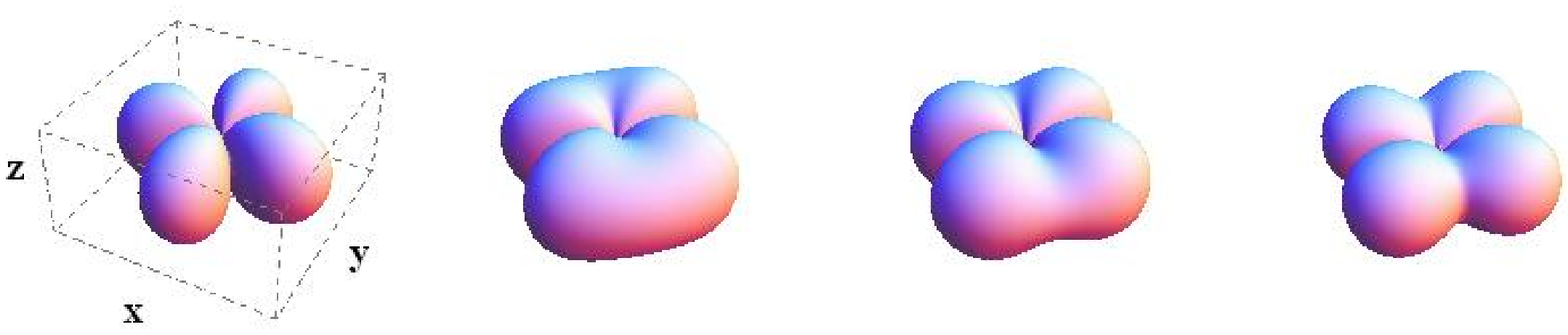}}
\subfigure{
\includegraphics[width=35pc]{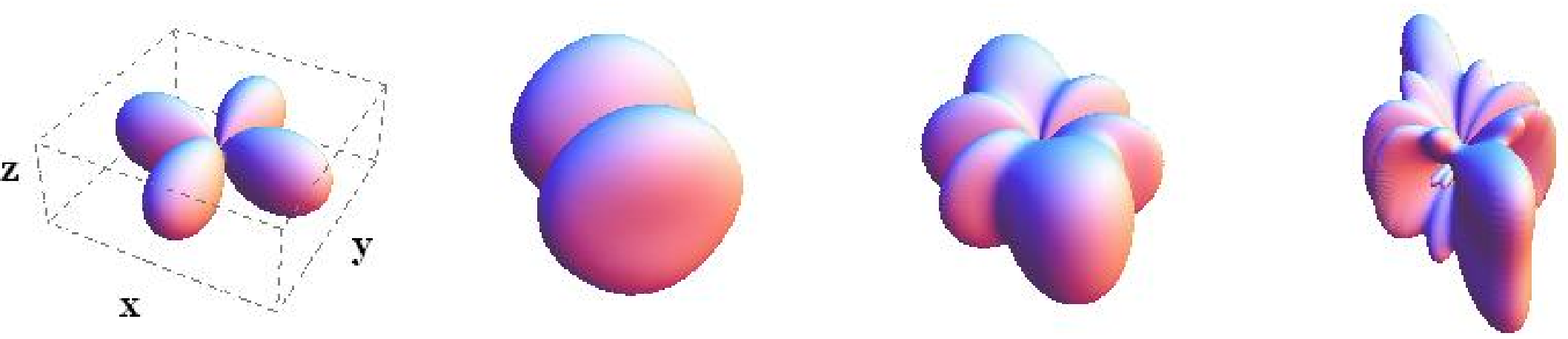}}
\subfigure{
\includegraphics[width=36pc]{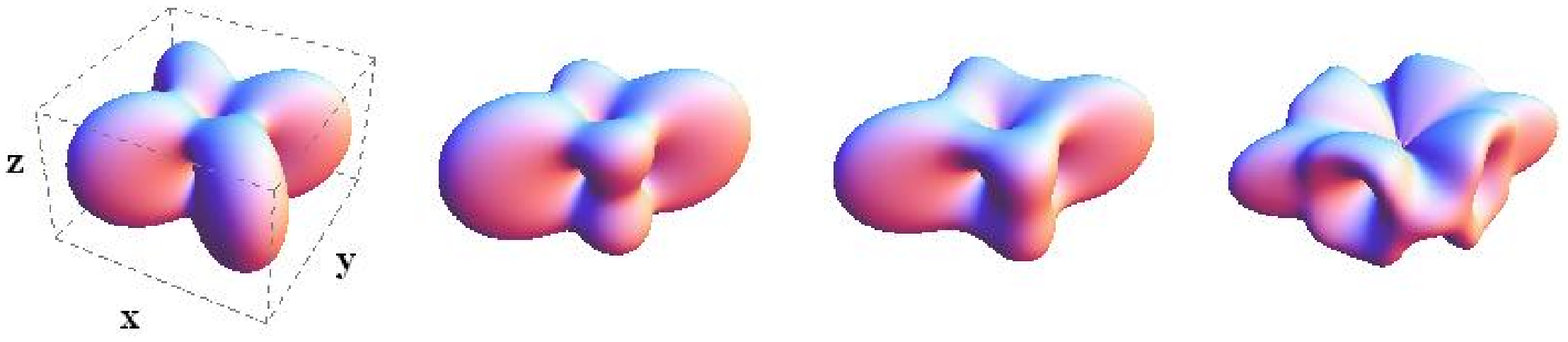}}
\subfigure{
\includegraphics[width=36pc]{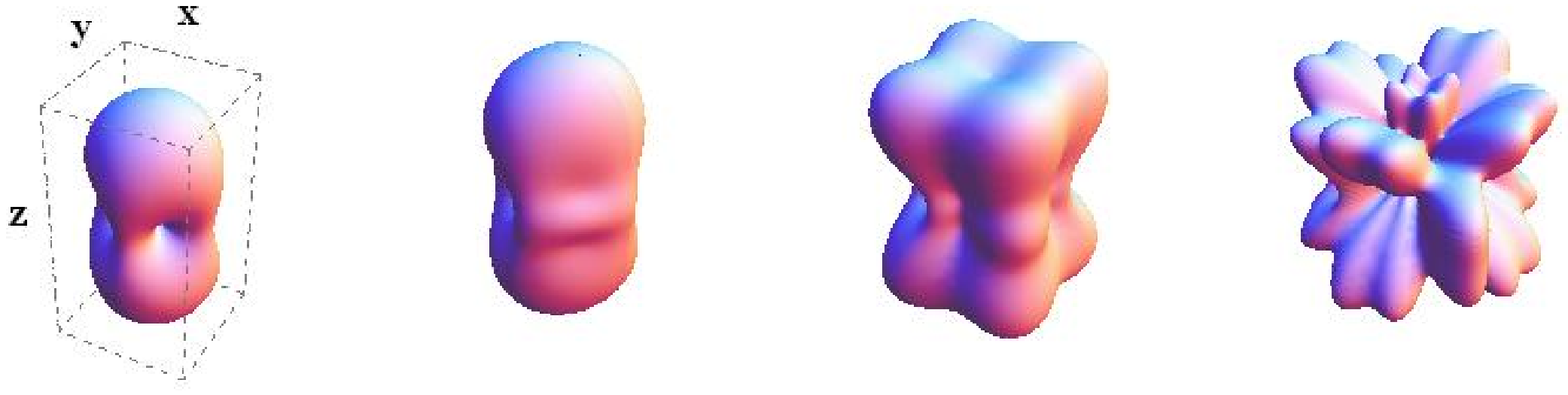}}
\end{center}
\caption{Shapes of the frequency response functions of the equal-arm
Michelson interferometer for different polarization mode.
From above: scalar longitudinal, scalar transversal,
vectorial and tensorial polarization.
We use the normalized frequencies $\omega L$, where $L$ is the arm length.
From the left: $\omega L=0.1,\;2,\;5$ and $10$. The scale is not preserved.
\label{f:shape:pi}}
\end{figure}

\begin{figure}[htp]
\begin{center}
\subfigure[][]{
\label{f:a}
\includegraphics[width=12pc]{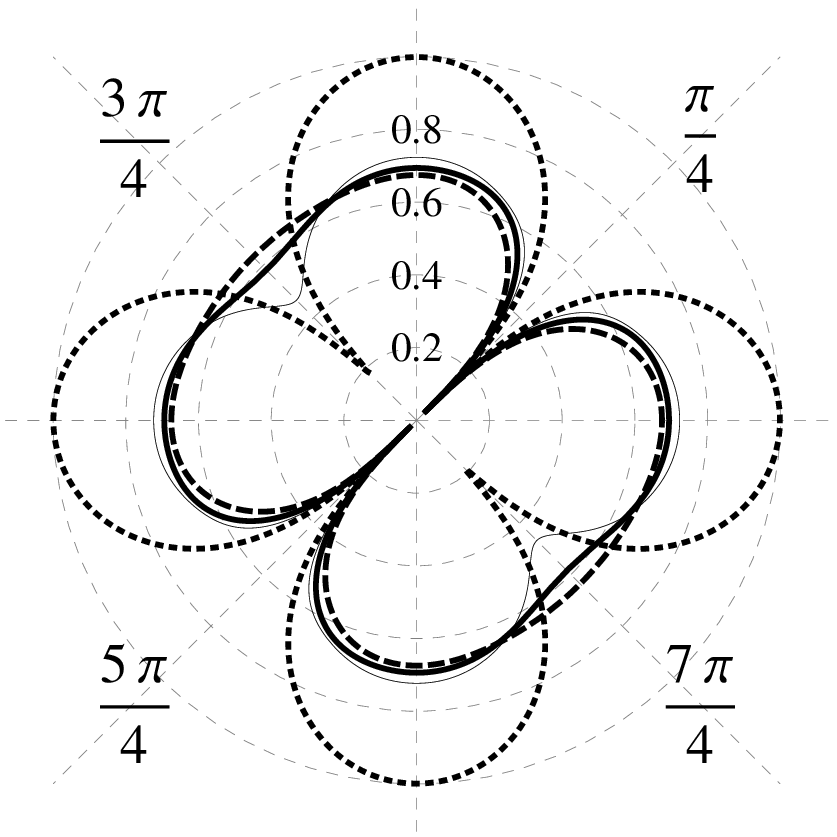}}
\qquad\subfigure[][]{
\label{f:b}
\includegraphics[width=12pc]{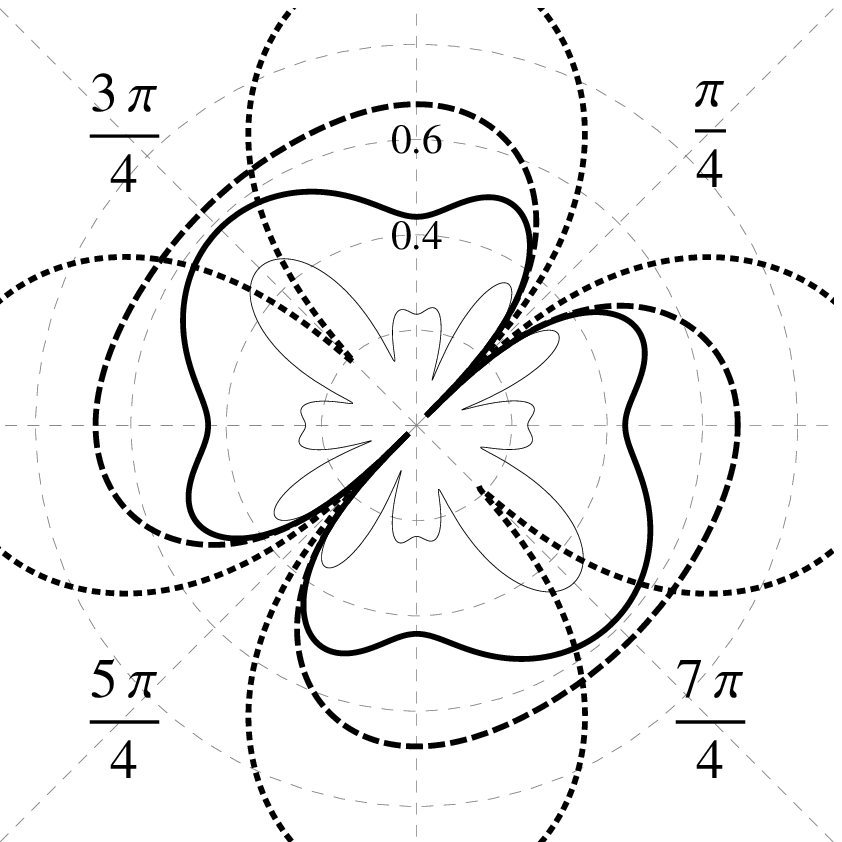}}
\hspace{1cm}

\subfigure[][]{
\label{f:c}
\includegraphics[width=12pc]{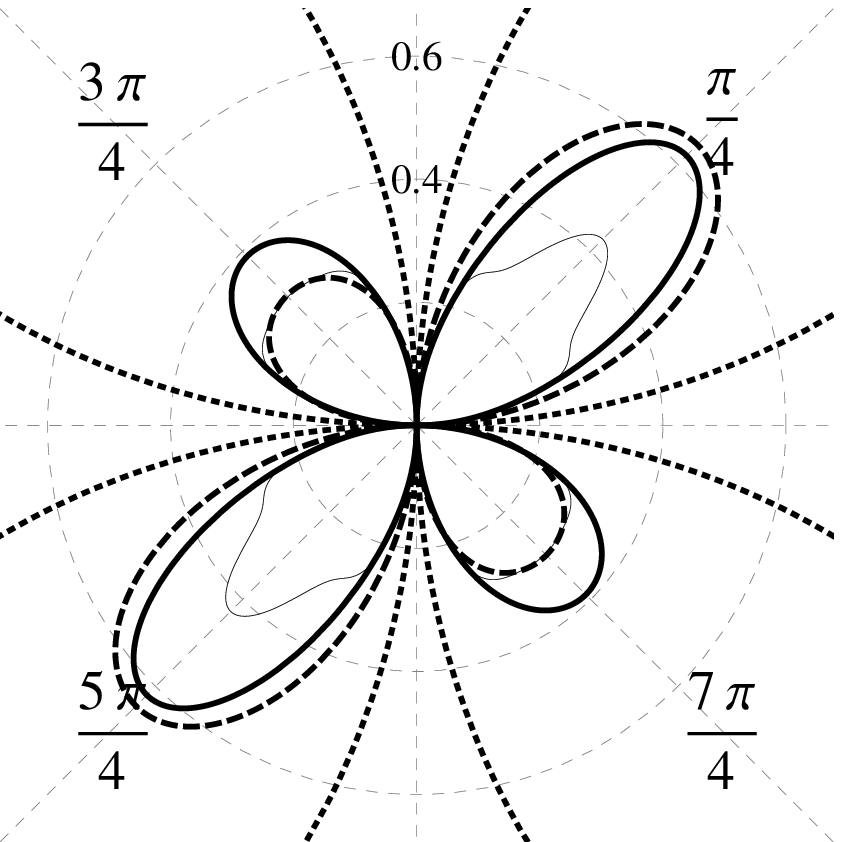}}
\qquad\subfigure[][]{
\label{f:d}
\includegraphics[width=12pc]{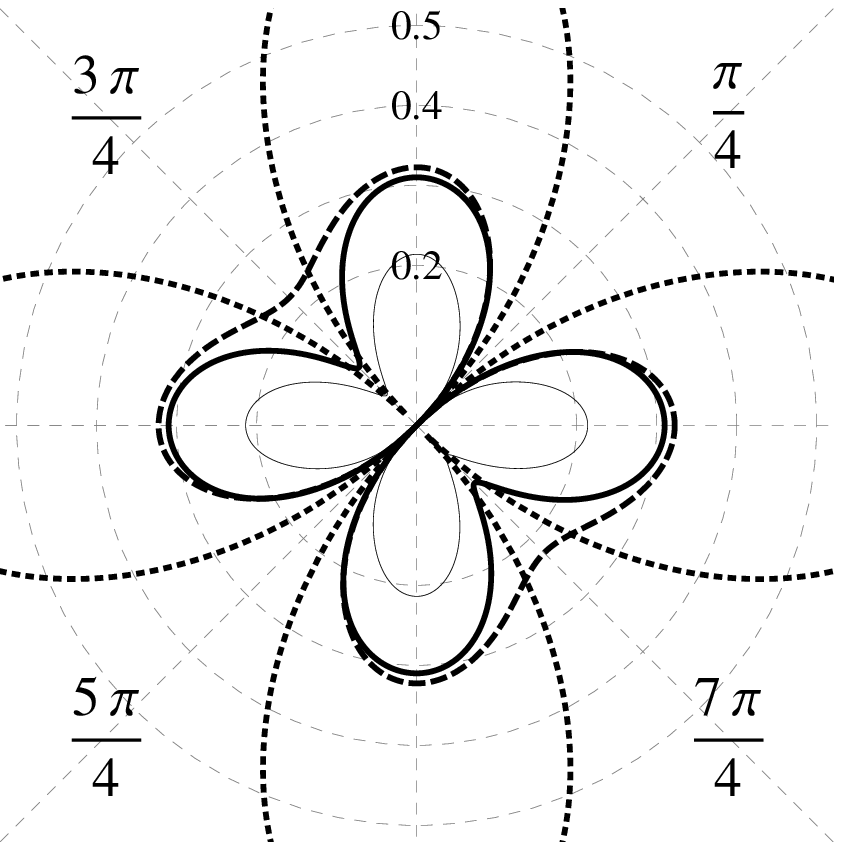}}
\hspace{1cm}

\subfigure[][]{
\label{f:e}
\includegraphics[width=12pc]{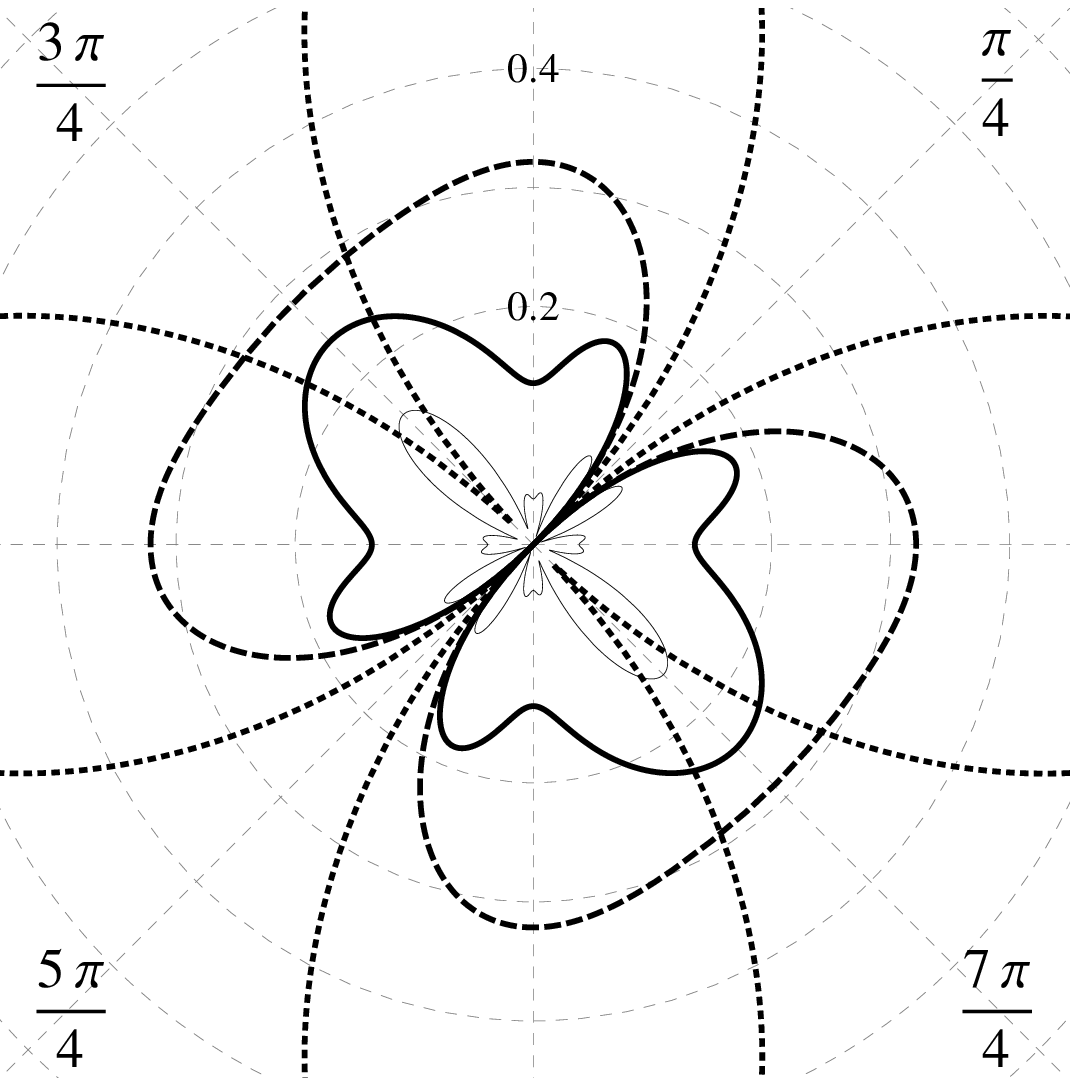}}
\qquad\subfigure[][]{
\label{f:f}
\includegraphics[width=12pc]{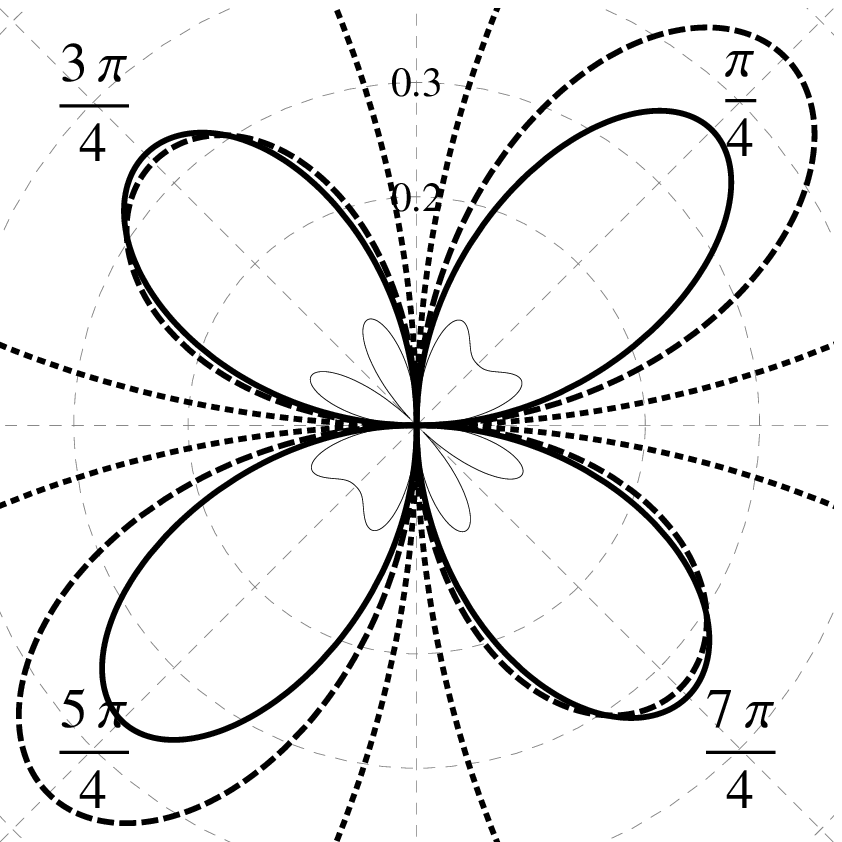}}
\end{center}
\caption{
Frequency response $|H_{M}(\omega)|$ of the equal-arm Michelson interferometer
as the function of the azimuthal angle $\phi$, fixed polar angle $\theta$ and $\psi=0$ for
scalar longitudinal ($\theta=\pi/2$) \subref{f:a},
scalar transversal ($\theta=\pi/2$) \subref{f:b},
vectorial $x$ ($\theta=\pi/2$) \subref{f:c},
vectorial $y$ ($\theta=\pi/4$) \subref{f:d},
tensorial $+$ ($\theta=\pi/2$) \subref{f:e} and
tensorial $\times$ ($\theta=\pi/4$) \subref{f:f} polarization.
Dotted, dashed, thick and thin curves correspond to the
normalize angular frequencies $x$ equal $0.1$, $2$, $5$ and $10$ respectively.
The frequency response has units $[L]$.
\label{f:angular:pi}}
\end{figure}

In Figs \ref{f:shape:pi} and \ref{f:angular:pi} we give plots of the frequency
responses of the equal-arm Michelson interferometer in the configuration shown in
Fig.\ref{f:frame2} as functions of the sky position and as functions of the azimuthal
angle for sections $\theta=const.$ for some selected frequencies.

In Fig. \ref{f:shape:pi} for the vectorial and tensorial modes we take the sum of the squared responses
\begin{eqnarray}
\left[H_M^v(\omega)\right]^2 & = & \frac12(|H_M^{vx}(\omega;\theta,\phi,\psi)|^2 +
|H_M^{vy}(\omega;\theta,\phi,\psi)|^2) \\
\left[H_M^t(\omega)\right]^2 & = & \frac12(|H_M^{tp}(\omega;\theta,\phi,\psi)|^2 +
|H_M^{tc}(\omega;\theta,\phi,\psi)|^2)
\end{eqnarray}
which are polarization angle $\psi$-independent and are equal to the $\psi$-square-averaged
$H_M^{vx}$ and $H_M^{tp}$ responses (or, equally, square-averaged $H_M^{vy}$ and $H_M^{tc}$).

\begin{figure}[htp]
\begin{center}
\includegraphics[width=30pc]{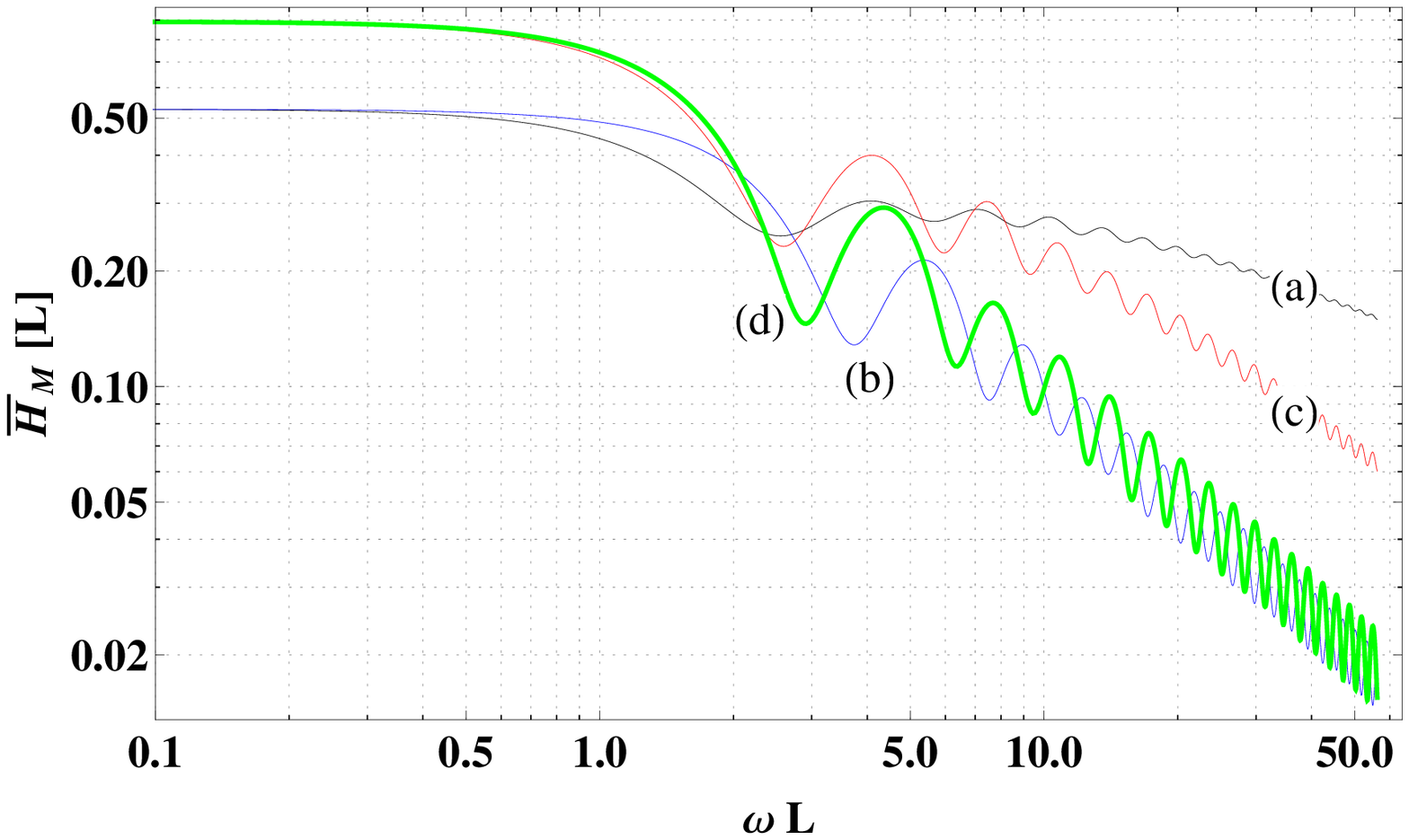}
\end{center}
\caption{The frequency responses $|H_M|$ of the equal-arm Michelson interferometer
averaged over the polarizations and sky location of sources as a function of $\omega L$:
(a) - scalar longitudinal, (b) - scalar transversal, (c) - vectorial
and (d) - tensorial polarization. The frequency response has units $[L]$.
To obtain plots of $H_{DLM}$ and $H_{FPM}$ one should multiply $|H_M(\omega)|$ by
$|\frac{\sin{N\omega L}}{\sin{\omega L}}|$ and $|\frac{1+\rho}{1-\rho e^{-2 i \omega L}}|$
respectively.}
\label{f:aver_resp:f}
\end{figure}

In Fig.\ref{f:aver_resp:f} we plotted the frequency responses averaged over the polarization
and position of the source as functions of the frequency. To compute the rms responses
we perform the Monte Carlo averaging of $H_{M,sl}$, $H_{M,st}$,
$H_{M,v}$ and $H_{M,t}$ over the sky location assuming uniform distribution of sources;
we take $10^4$ points for each value of the frequency. The low-frequency level
of the sensitivities is determined by the angular pattern functions $F^{\pi}({\bf n}_1)-F^{\pi}({\bf n}_2)$;
approximately for frequencies $\omega L>5$ the responses $|H_{M}|(\omega)$ fall as $1/(\omega L)$ for all
modes except the scalar longitudinal mode in which case it falls as $1/\sqrt{\omega L}$.

%

\subsection{Triangular configurations. Correlation analysis}
\label{ch:triangular}

To exemplify the results obtained in the previous sections we apply the exact
formulas in the numerical analysis of the antenna pattern functions for
the stochastic gravitational-wave signal.

\begin{figure}[htp]
\begin{center}
\includegraphics[width=18pc]{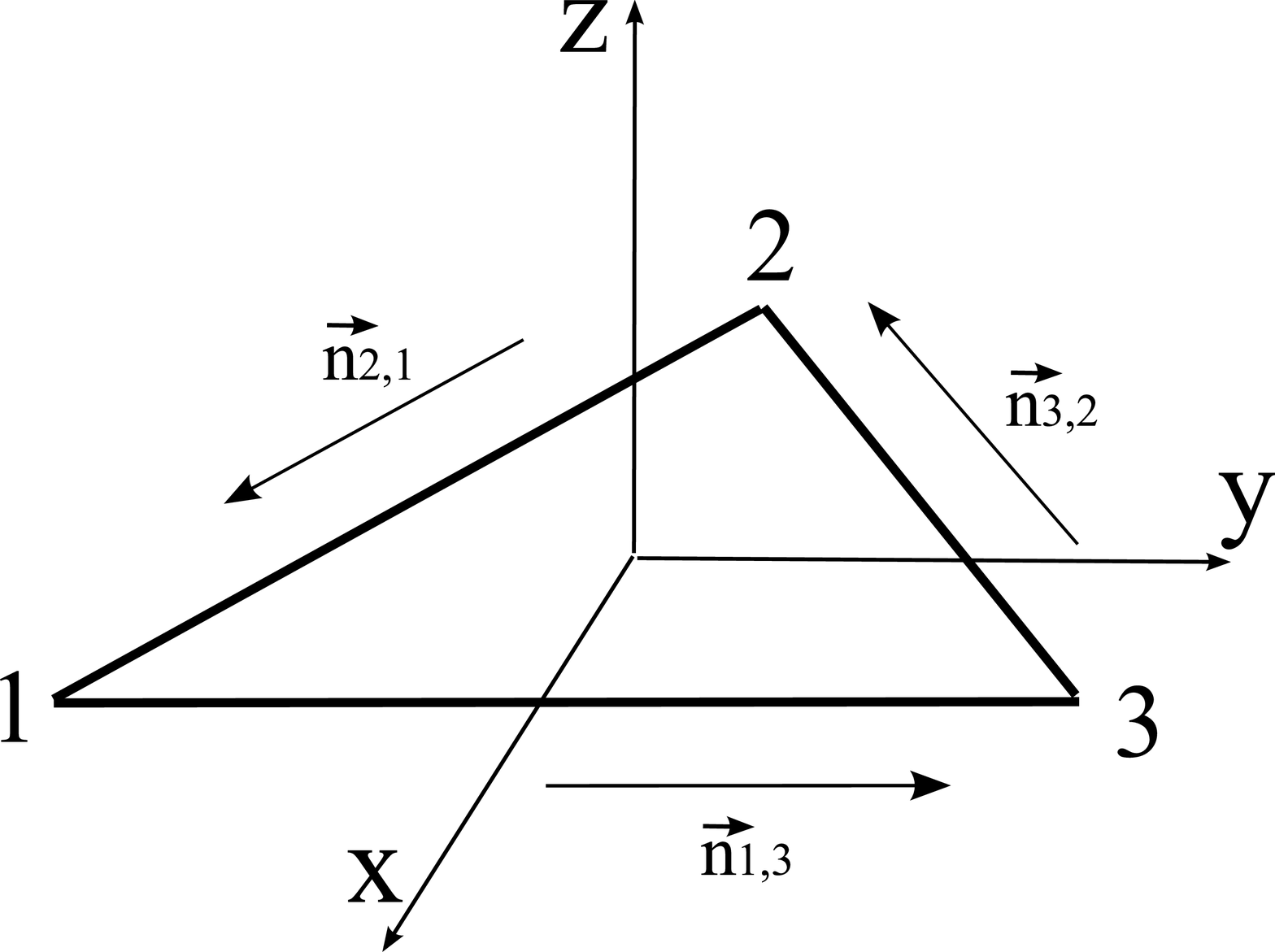}
\end{center}
\caption{
Triangular configuration of spacecraft.
\label{f:lisa1}}
\end{figure}

Most planned space interferometers (LISA, BBO, DECIGO) consist of one or more spacecraft
constellations forming almost equilateral triangles, Fig. \ref{f:lisa1}.
The exact configurations and orbital parameters of spacecraft are still debated and presumably
detectors will have different interferometric designs. LISA, BBO and ultimate-DECIGO
(an ultimate version of DECIGO whose sensitivity is limited only by the quantum noise)
will be transponder-like interferometers and will use the so called time delay interferometry
\cite{Armstrong99} while the recently proposed DECIGO interferometer will be equipped with
Fabry-Perot cavities \cite{Kawamura08}. Among the principal targets of the space interferometers
are the stochastic signals produced by the unresolved population of the Galactic and extragalactic
binaries and the primordial gravitational-wave background. The astrophysical foreground is expected
to form an inhomogeneous signal whose spatial fluctuations shall trace the density distribution of
sources. Thus to make a sky-maps of the background one should explore its anisotropic components.
For this reason the sensitivity of detectors for different multipole moments was studied
in \cite{Cornish02},\cite{Kudoh05} and \cite{SetoCooray04}. On the other hand the primordial
gravitational-wave background has originated in a high-energy regime so one can speculate
that the additional polarization components may contribute in a significant way. Thus the detection
and discrimination of various polarization components would give an insight in the physics of
the early Universe \cite{Nishizawa10}. Here we give the angular and frequency
characteristic of the triangular detector for the stochastic signals in different polarization states.
We use the spherical harmonics analysis to compute the lowest multipole moments (up to $l=6$)
of the antenna pattern function in the static spacecraft configuration for the self and
cross-correlated signals which are free from the noise correlations.

We consider the detector configuration from Fig. \ref{f:lisa1}.
For the equal-arm Michelson interferometer receiving a signal at the spacecraft $1$
the detector tensor is given by
\begin{eqnarray}
\label{eq:M1dten}
{\bf D}_{M_1}(\omega,{\bf\Omega}) & = & \frac{L}{2}\left\{
{\bf n}_{1,3}\otimes{\bf n}_{1,3}
\left[
{\cal T}(\omega;L{\bf n}_{3,1},{\bf \Omega})+{\cal T}(\omega;L{\bf n}_{1,3},{\bf \Omega})
e^{-i\omega\,L(1+{\bf\Omega}\cdot{\bf n}_{1,3})}
\right]-\right.\nonumber\\
&&
\left.
{\bf n}_{1,2}\otimes{\bf n}_{1,2}
\left[
{\cal T}(\omega;L{\bf n}_{2,1},{\bf \Omega})+{\cal T}(\omega;L{\bf n}_{1,2},{\bf \Omega})
e^{-i\omega\,L(1+{\bf\Omega}\cdot{\bf n}_{1,2})}]
\right]
\right\}\nonumber\\
& = &
L\,
{\bf n}_{1,3}\otimes{\bf n}_{1,3}\,{\cal T}_{1}(\omega;L{\bf n}_{1,3},{\bf \Omega})
-L\,{\bf n}_{1,2}\otimes{\bf n}_{1,2}\,{\cal T}_{1}(\omega;L{\bf n}_{1,2},{\bf \Omega})
\end{eqnarray}
and we have
\begin{eqnarray}
\label{eq:M1}
{\bf D}_{M_1}(\omega,{\bf\Omega}):{\boldsymbol \epsilon}^{\pi}({\bf\Omega}) & = &
2\,L\,{\cal T}_1(\omega;L{\bf n}_{1,3},{\bf\Omega})F^{\pi}({\bf n}_{1,3}) -
2\,L\,{\cal T}_1(\omega;L{\bf n}_{1,2},{\bf\Omega})F^{\pi}({\bf n}_{1,2});
\end{eqnarray}
signals $M_2$ and $M_3$ are defined similarly. The unequal-arm Michelson combinations
$X$, $Y$, $Z$ are given by $X(t)=M_{1}(t)-M_{1}(t-2L),\;$etc. According to Eq.(\ref{eq:dten})
the detector tensors define the antenna pattern functions ${\cal F}_{M_IM_J}$.
Given ${\cal F}_{M_IM_J}$ one obtains the antenna pattern functions of the
Fabry-Perot-Michelson interferometer,
${\cal F}_{FPM_IFPM_J}=\frac{(1+\rho)^2}{(1-\rho)^2 + 4\rho \sin^2{\omega L}}{\cal F}_{M_IM_J}$,
and of the unequal-arm Michelson interferometer,
${\cal F}_{X_IX_J}=4\sin^2{(\omega L)}\,{\cal F}_{M_IM_J}$.

\begin{figure}[htp]
\begin{center}
\renewcommand{\arraystretch}{1.2}
\subfigure{
\includegraphics[width=18pc]{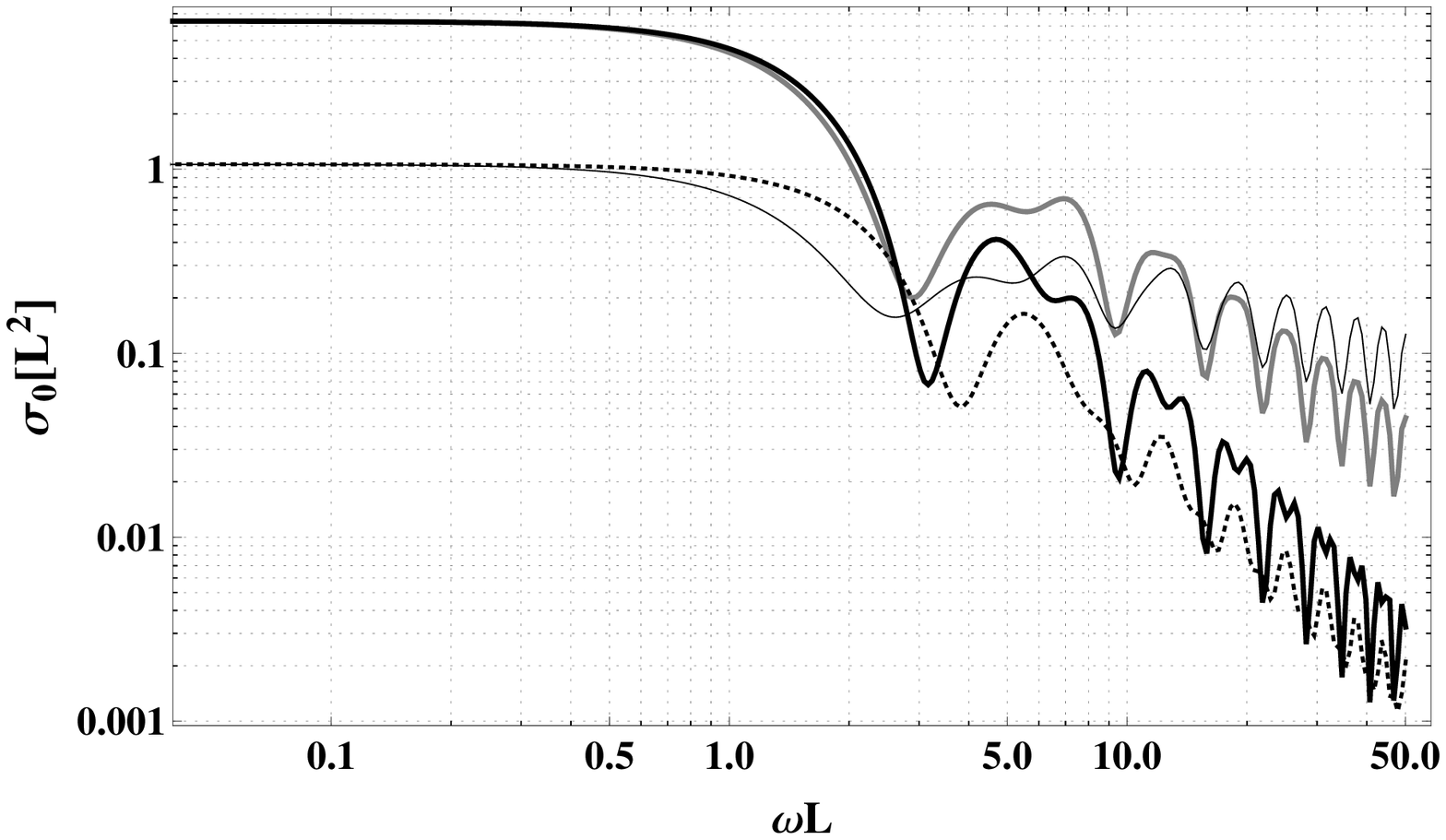}}
\qquad\subfigure{
\includegraphics[width=18pc]{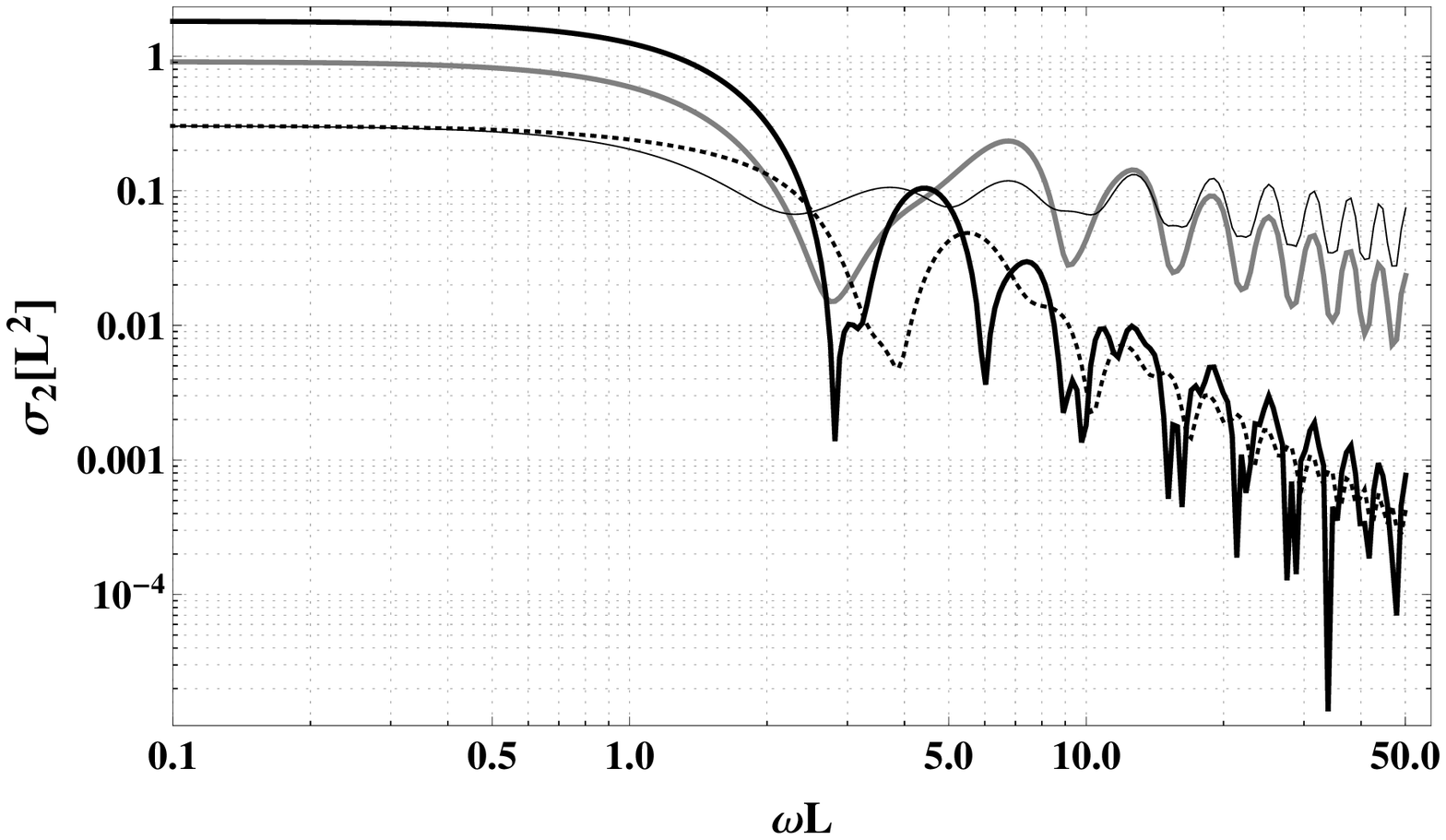}}
\subfigure{
\includegraphics[width=18pc]{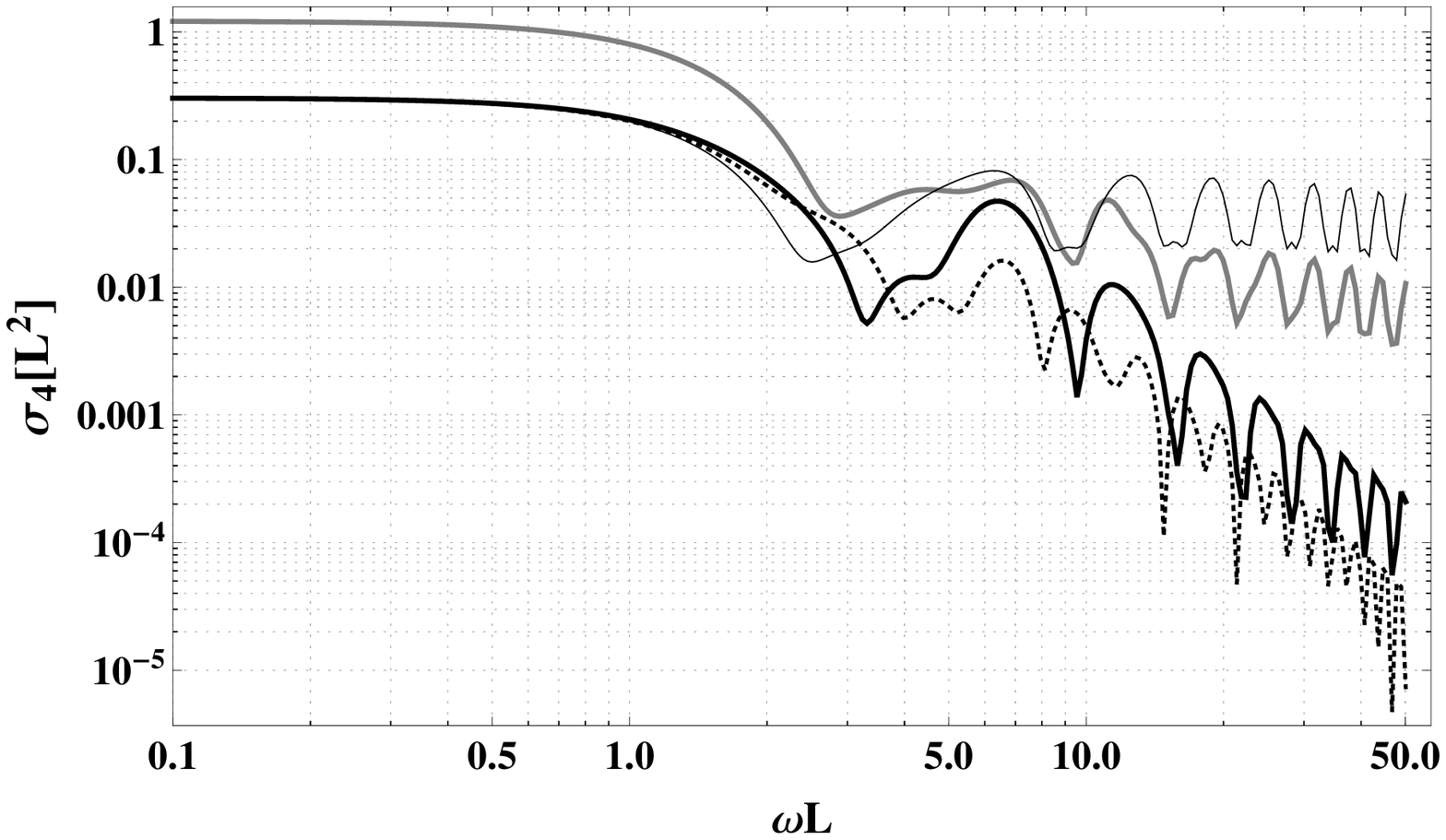}}
\qquad\subfigure{
\includegraphics[width=18pc]{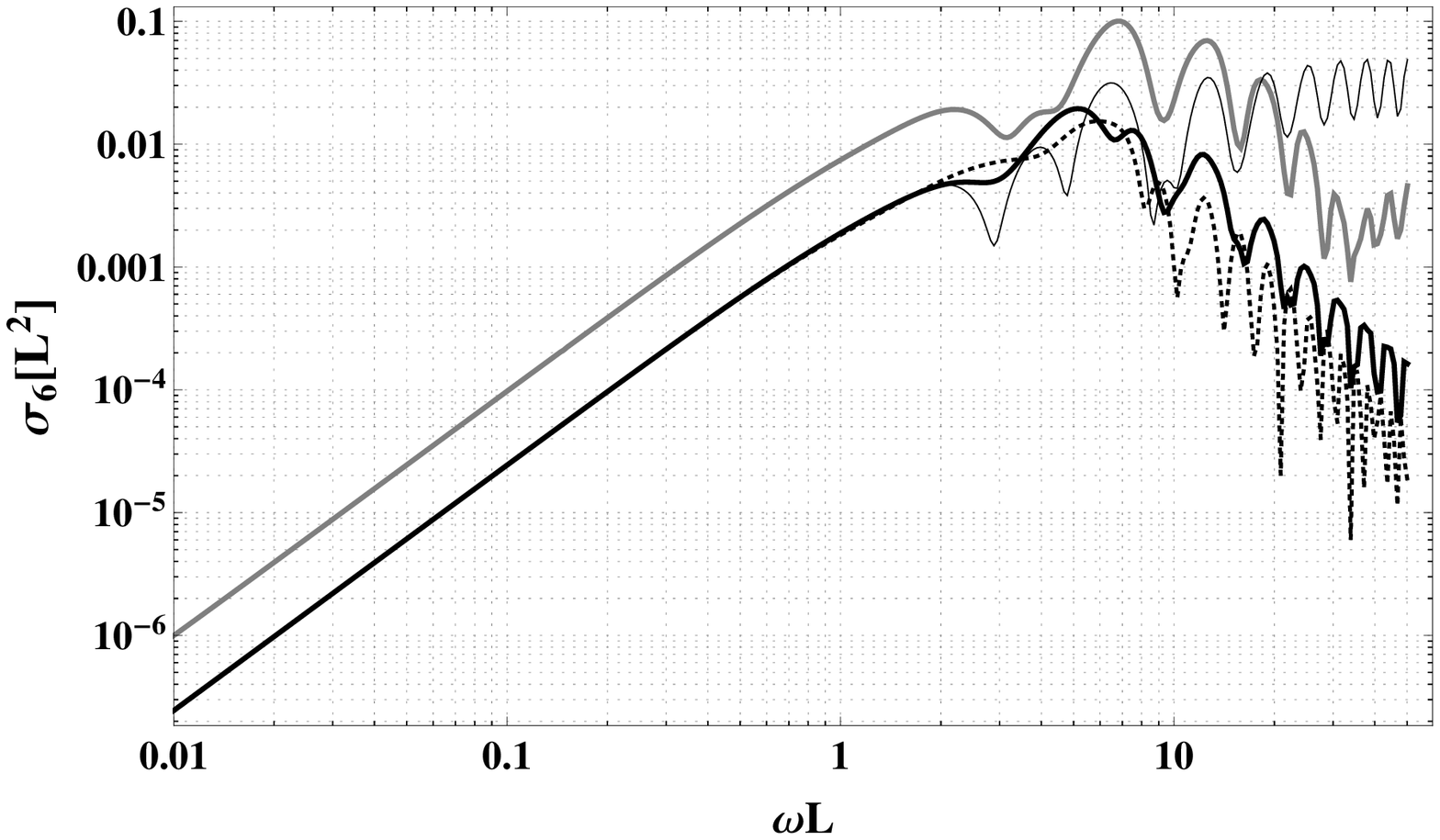}}
\end{center}
\caption{
Self-correlations of the optimal $A_M$ observable as a function of the frequency
for $l=0$, $2$, $4$ and $6$ for the tensorial (thick black curve), vectorial (thick gray curve),
scalar longitudinal (thin curve) and scalar transversal (thin dotted curve) polarization.
The angular power has units $[L^2]$.
\label{f:self:l06}}
\end{figure}

\begin{figure}[htp]
\begin{center}
\subfigure{
\includegraphics[width=18pc]{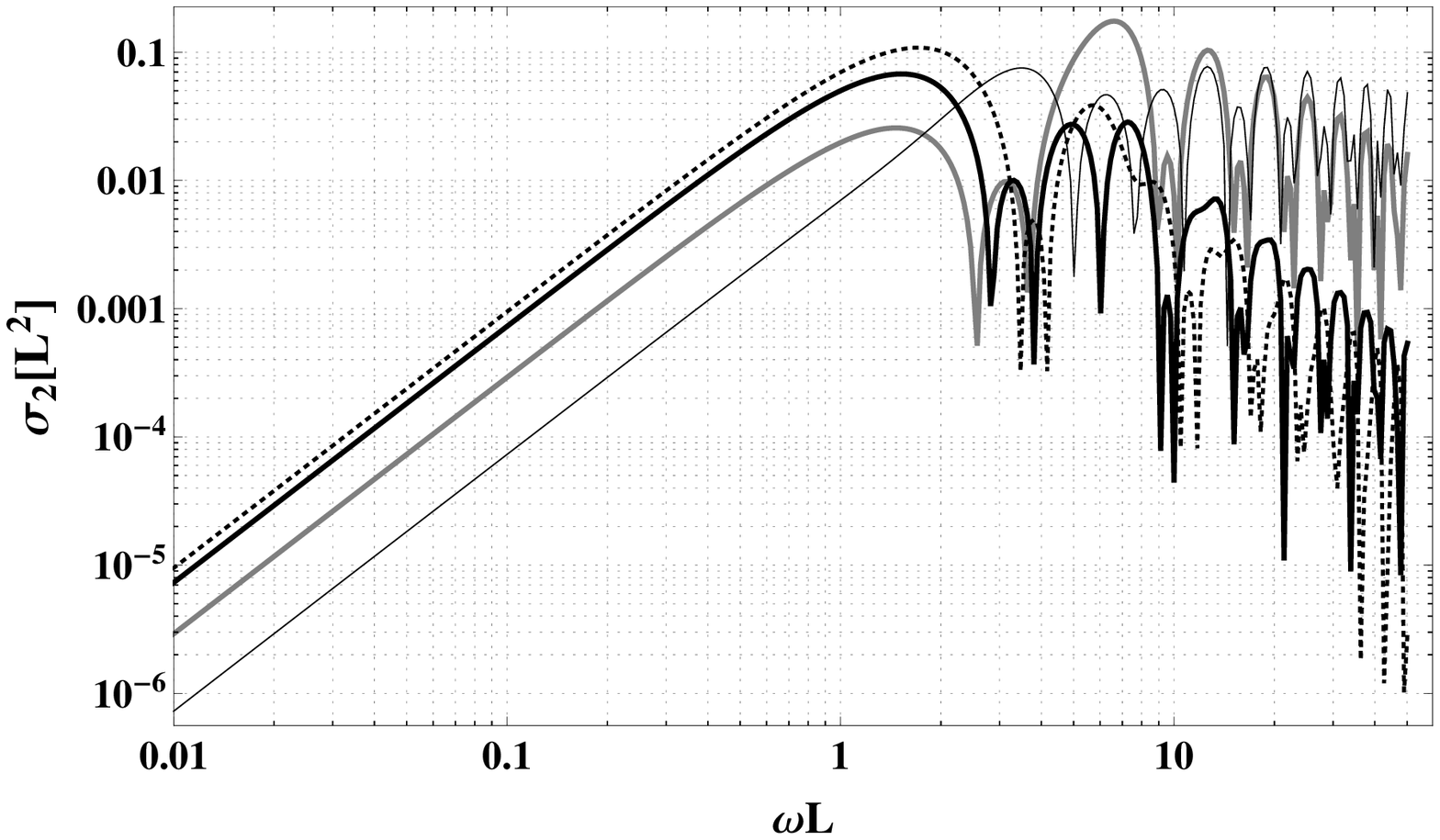}}
\qquad\subfigure{
\includegraphics[width=18pc]{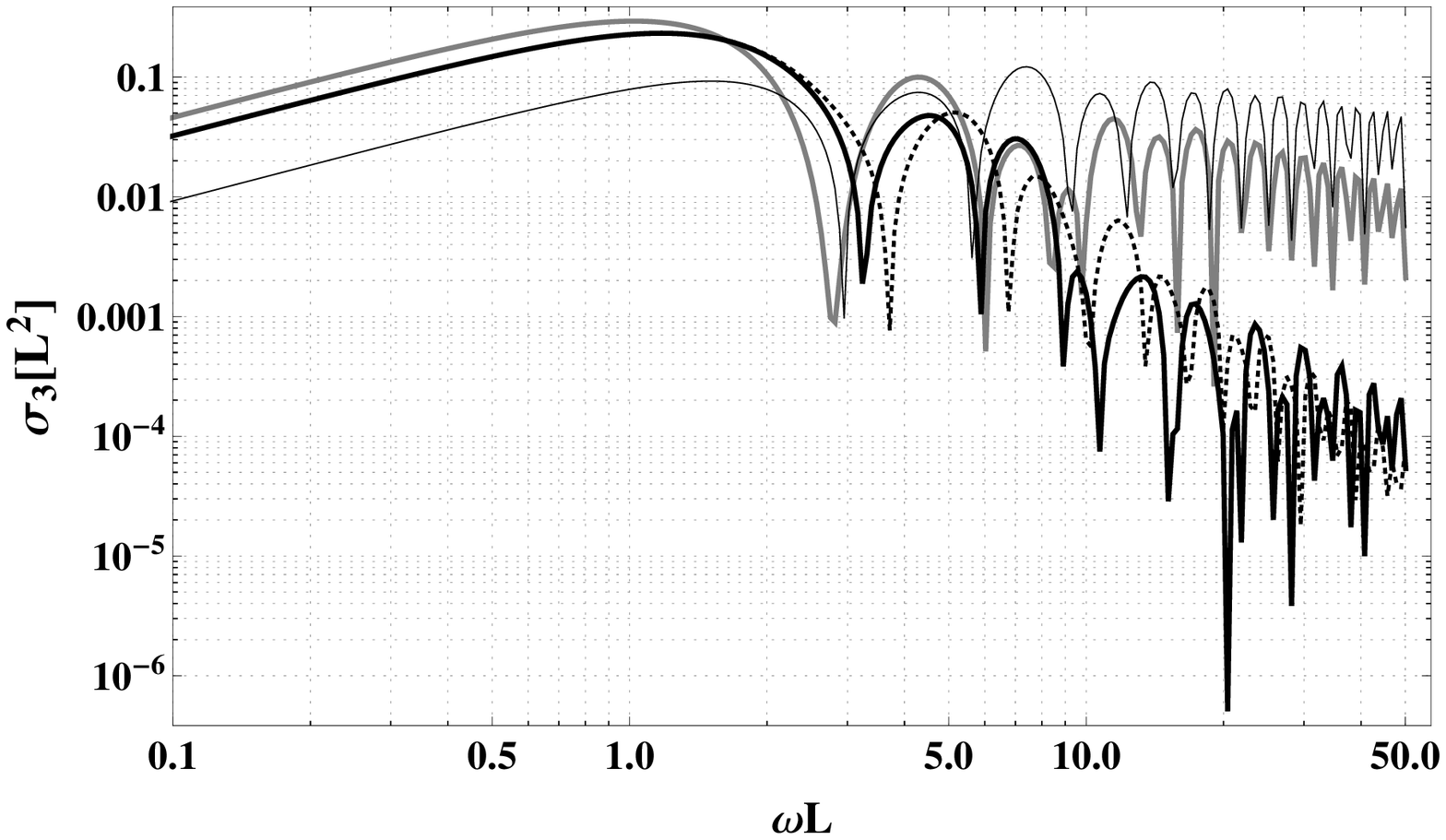}}
\subfigure{
\includegraphics[width=18pc]{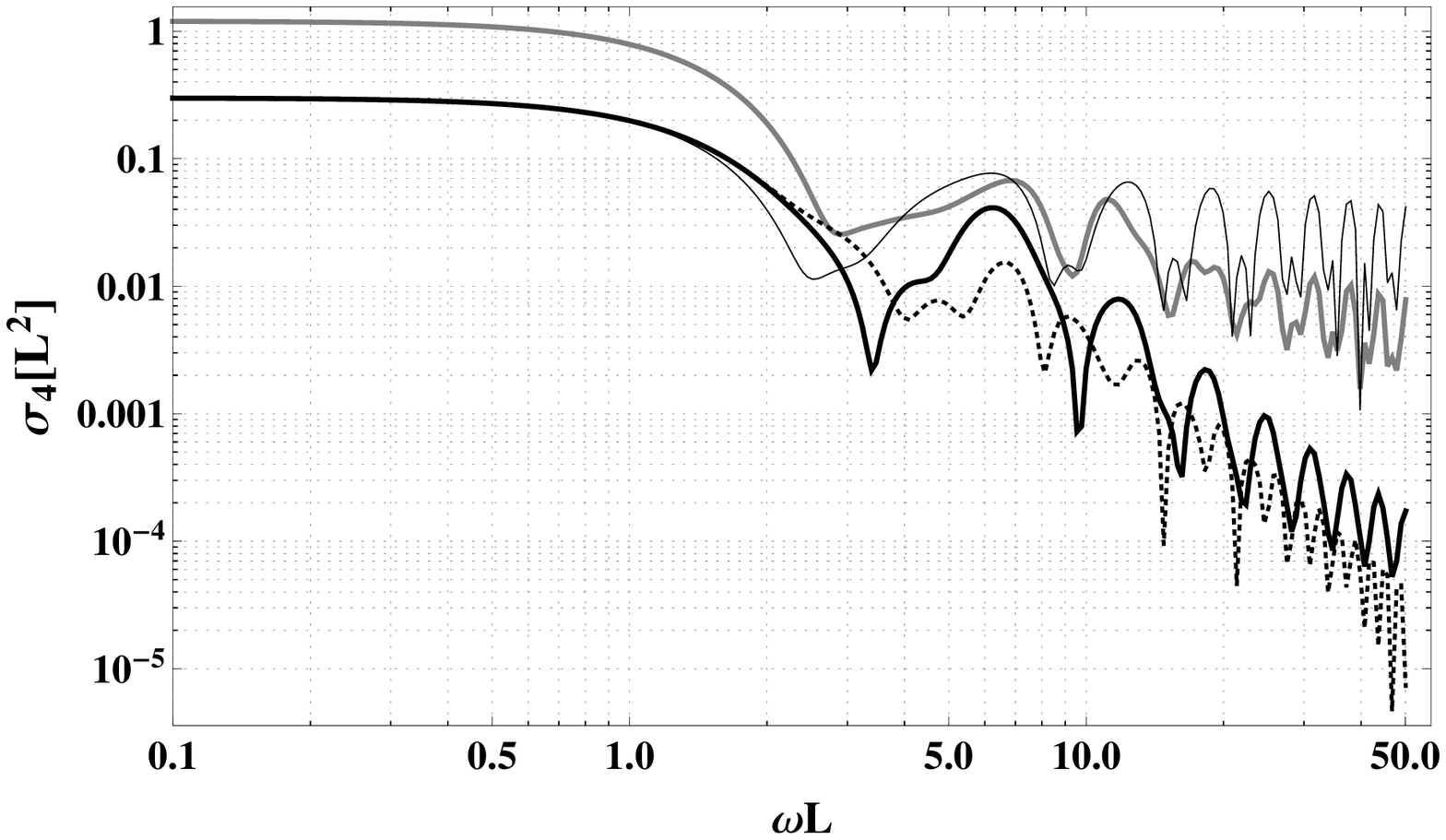}}
\qquad\subfigure{
\includegraphics[width=18pc]{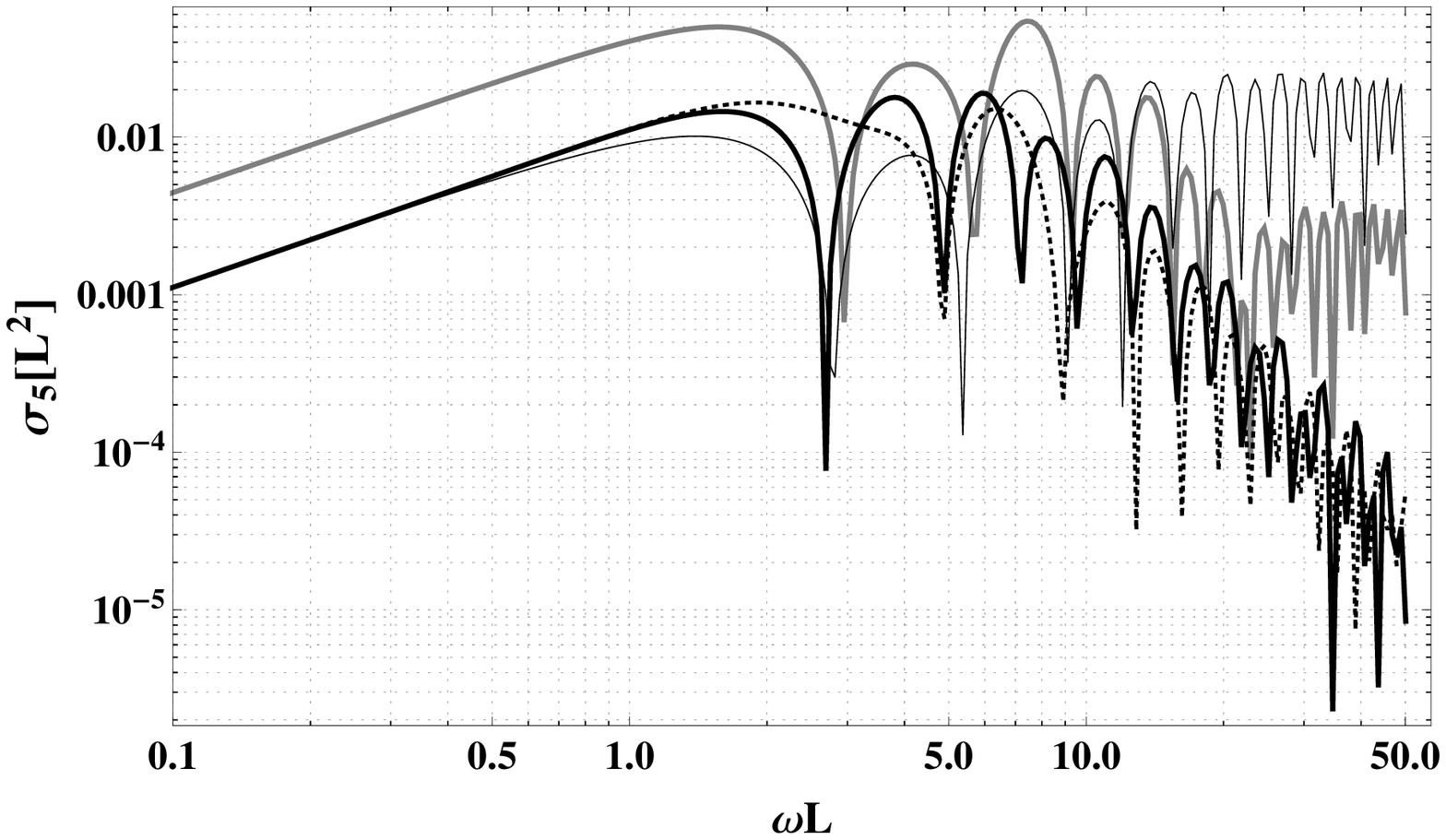}}
\subfigure{
\includegraphics[width=18pc]{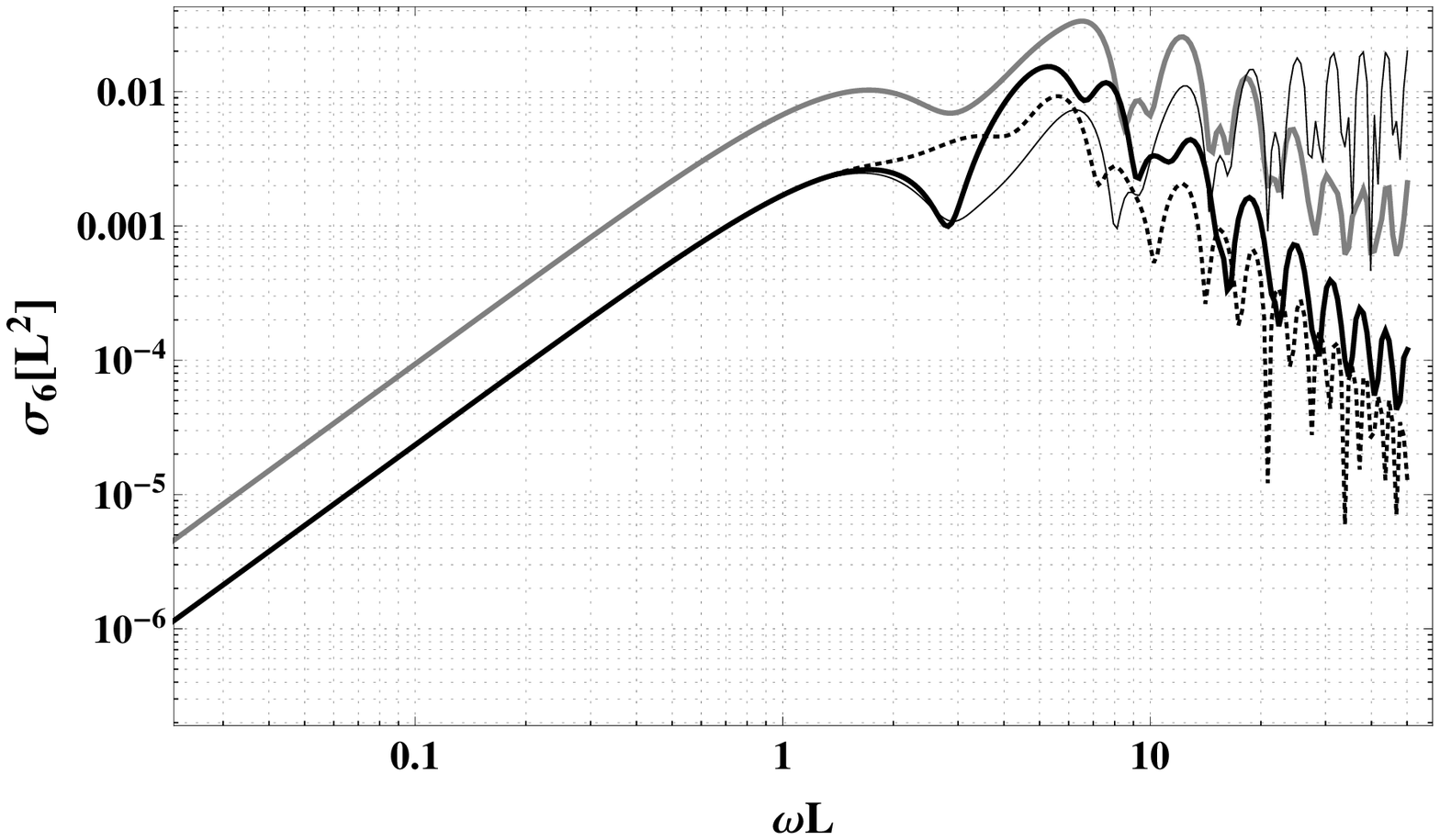}}
\end{center}
\caption{
Cross-correlations of the optimal $A_M$ and $E_M$ observables as a function of the frequency
for $l=2$, $3$, $4$ $5$ and $6$. Different polarization modes are marked as in Fig. \ref{f:self:l06}.
\label{f:cross:l26}}
\end{figure}

In the cross-correlation analysis of two weak signals the covariance matrix of noises should
be diagonal. To illustrate the properties of the angular patterns we restrict in this section
to the two noise-orthogonal combinations constructed from $M_I$,
\begin{equation}
A_M = \frac{1}{\sqrt{2}}\left(-M_1 + M_3\right),\qquad
E_M = \frac{1}{\sqrt{6}}\left(M_1 - 2M_2 + M_3\right).
\end{equation}
The optimal $A$, $E$ observables are analogous combinations of $X$, $Y$, $Z$ so that their
antenna pattern functions are given by
${\cal F}_{AA}=4\sin^2{(\omega L)}\,{\cal F}_{A_MA_M}$ and
${\cal F}_{AE}=4\sin^2{(\omega L)}\,{\cal F}_{A_ME_M}$.
We decompose the antenna patterns with respect to the spherical harmonics
\begin{equation}
a_{lm}(\omega L)=\int_{S^2}Y_{lm}^{*}({\bf\Omega}){\cal F}(\omega L,{\bf\Omega})\frac{d{}\bf\Omega}{4\pi}
\end{equation}
and define the rotation-invariant quantity that characterizes the contribution
of the $l$-th multipole mode to the angular power of the antenna pattern function
\begin{equation}
\sigma_{l}^{2}(\omega L)=\frac{1}{2l+1}\sum_{m=-l}^{l}|a_{lm}(\omega L)|^2.
\end{equation}
Figs. \ref{f:self:l06} and \ref{f:cross:l26} show the dependence of the angular powers
on the frequency for all polarization modes for the self-correlated $A_M$ and
cross-correlated $A_M$, $E_M$ optimal observables. The angular power for the optimal
$A$ and $E$ responses can be obtained by multiplying $\sigma_l^{A_MA_M}$ and $\sigma_l^{A_ME_M}$
by $4\sin^2{\omega L}$.

\begin{table}
\begin{center}
\renewcommand{\arraystretch}{1.3}
\begin{tabular}{c c c c c l}
\hline\hline
\multirow{1}{1.2cm}{$\sigma_l$}
&
\multicolumn{1}{p{2.3cm}}
{\centering scalar l }
&
\multicolumn{1}{p{2.3cm}}
{\centering scalar t }
&
\multicolumn{1}{p{2.3cm}}
{\centering vector }
&
\multicolumn{1}{p{2.3cm}}
{\centering tensor }
&
\multicolumn{1}{p{2.3cm}}
{\centering  }
\\
\hline
$\sigma_0$
&
$\frac16$ & $\frac16$ & $1$ & $1$ &
$\times\frac{18 \sqrt{\pi}}{5}$

\\
$\sigma_2$
&
$\frac16$ & $\frac16$ & $\frac12$ & $1$ &
$\times\frac{36 \sqrt{\pi}}{35}$
\\
$\sigma_4$
&
$1$ & $1$ & $4$ & $1$ &
$\times\frac{6 \sqrt{\pi}}{35}$
\\
$\sigma_6$
&
$1$ & $1$ & $4$ & $1$
& $\times\frac{\sqrt{\pi}}{8008}\sqrt{\frac{1829}{15}}(\omega L)^2$
\\
\hline\hline
\end{tabular}
\end{center}
\caption{The leading terms in the long-wavelength limits of the angular power $\sigma_l$
for the self-correlated optimal $A_M$ observables.}
\label{t:selfLW}
\end{table}

\begin{table}
\begin{center}
\renewcommand{\arraystretch}{1.3}
\begin{tabular}{c c c c c l}
\hline\hline
\multirow{1}{1.5cm}{$\sigma_l$}
&
\multicolumn{1}{p{2.3cm}}
{\centering scalar l }
&
\multicolumn{1}{p{2.3cm}}
{\centering scalar t }
&
\multicolumn{1}{p{2.3cm}}
{\centering vector }
&
\multicolumn{1}{p{2.3cm}}
{\centering tensor }
&
\multicolumn{1}{p{2.3cm}}
{\centering  }
\\
\hline
$\sigma_2$
&
$\frac{1}{10}$ & $\frac{13}{10}$ & $\frac{2}{5}$ & $1$ &
$\times\frac{1}{14}\sqrt{\frac{\pi}{3}}(\omega L)^2$
\\
$\sigma_3$
&
$\frac{2}{7}$ & $1$ & $\frac{10}{7}$ & $1$ &
$\times\sqrt{\frac{\pi}{30}}\,\omega L$
\\
$\sigma_4$
&
$1$ & $1$ & $4$ & $1$ &
$\times\sqrt{\frac{\pi}{35}}$
\\
$\sigma_5$
&
$1$ & $1$ & $4$ & $1$
& $\times\frac{1}{11}\sqrt{\frac{\pi}{210}}\,\omega L$
\\
$\sigma_6$
&
$1$ & $1$ & $4$ & $1$
& $\times\frac{1}{52}\sqrt{\frac{\pi}{210}}(\omega L)^2$
\\
\hline\hline
\end{tabular}
\end{center}
\caption{The leading terms in the long-wavelength limits of the angular power $\sigma_l$
for the cross-correlated optimal $A_M$ and $E_M$ observables.}
\label{t:crossLW}
\end{table}
Since the spherical harmonics satisfy $Y_{lm}(-{\bf\Omega})=(-1)^{l}Y_{lm}({\bf\Omega})$
and the antenna patterns of the self-correlated signals $M_IM_I$, $A_MA_M$, $E_ME_M$, $AA$,
$EE$ are even functions, ${\cal F}(\omega L,-{\bf\Omega})={\cal F}(\omega L,{\bf\Omega})$,
the odd multipole moments vanish for all polarizations. Furthermore, following arguments
given in \cite{Kudoh05} which make use of some geometric relations between optimal combinations
and transformation properties of the antenna pattern functions one can show that
$\sigma^{E_ME_M}_l=\sigma^{A_MA_M}_l$ ($\sigma^{EE}_l=\sigma^{AA}_l$) for all $l$ and that
angular powers of cross-correlated optimal signals $A_M$, $E_M$ ($A$, $E$) for $l=0$ and $l=1$
vanish for all polarization modes.

Plots in Figs. \ref{f:self:l06} and \ref{f:cross:l26} show the same frequency dependence for all
polarizations in the low frequency domain. The results are summarized in Tabs. \ref{t:selfLW} and
\ref{t:crossLW} which give the leading terms in the LW limits for all considered
angular powers. We observe that the main contributions to the angular powers for the self-correlated
signals in low frequencies comes from the $l=0$, $2$ and $4$ multipole moments and for the
cross-correlated signals from $l=4$ moment. We also observe that the tensor mode has the highest
angular power for the monopole ($l=0$) moment while starting from $l\geq2$ the angular power of
the vector mode dominates. For higher frequencies we notice $f^{-2}$ decay of the transversal
polarizations and slower decay of both longitudinal polarizations, the behavior analogous
to the sky-averaged frequency response depicted in the Fig. \ref{f:aver_resp:f}.

\subsection{More arms}
In this section we analyze detector's transfer functions for some multi-arm configurations.
They are interesting since various geometrical transfer functions differently affect
gravitational waves having different polarizations and also give an opportunity to distinguish
between the gravitational wave signals and instrumental noise. Particular importance
have configurations with the null signal transfer function. For example in the case of LISA space
interferometer various combinations of laser signals exchanged between spacecraft were proposed
which have null transfer function for gravitational wave signals in the long wavelength limit.
They were successfully used in \cite{Tinto01}, \cite{Cor10} and \cite{Robinson08} in the data analysis
of the simulated stochastic signals of Galactic and cosmological origins. We remark here that this
property is generally valid for waves having arbitrary polarization so one could utilize it to
compare and complement the correlation analysis proposed by Nishizawa et.al. \cite{Nishizawa10},
\cite{Nishizawa09} in the study of interferometric detection of various polarization modes in the
cosmological stochastic background.

In the recent paper \cite{TA10} it was noticed that in LISA detector scalar transversal
polarization mode has vanishing sensitivity for the frequencies equal to integer multiples of
the inverse of the one-way-light-time for the Sagnac $\alpha$ and symmetrized Sagnac $\zeta$
combinations in the approximation of equal arm lengths. We show here that this property holds
even in more general configurations, namely when the response is formed as a difference of
two round-trip signals taken in the opposite directions in a multi-arm interferometer.
It is assumed that the trajectory of photons forms a closed loop but otherwise the geometry
of paths and the number $N$ of links is arbitrary; the time delay in each link is an arbitrary
multiple of $L$. That is we show the vanishing sensitivity of the response
\begin{eqnarray}
\label{eq:multi}
&&y_{1,2}(t+m_{1}L) + y_{2,3}(t+m_{2}L) + \cdots + y_{N-1,N}(t+m_{N-1}L) + y_{N,1}(t+m_{N}L) - \\
&&[y_{1,N}(t+m'_{N}L) + y_{N,N-1}(t+m'_{N-1}L) + \cdots y_{3,2}(t+m'_{2}L) + y_{2,1}(t+m'_{1}L)]\nonumber
\end{eqnarray}
to a plane monochromatic wave
${\rm h}(t-{\bf\Omega}\cdot{\bf x}){\boldsymbol \epsilon}^{st}=e^{i\omega(t-{\bf\Omega}\cdot{\bf x})}{\boldsymbol \epsilon}^{st}$
 with the frequency $\omega=2\pi n/L$, for a round trip marked by points $1,2,\ldots,N-1,N,1$
and with arbitrary integers $m_{k},\,m'_{k}$, $\;k=1,2,\ldots N$.
To see this we note that according to Eqs.(\ref{eq:yab}) and (\ref{eq:uv:st}) the contribution
to the response (\ref{eq:multi}) coming from the $k$-th link oriented along the unit vector
${\bf n}_{k,k+1}$  is given by
\begin{eqnarray}
\label{eq:st:nk}
y_{k,k+1}(t + m_k L) & = &
\frac12 \left[
{\rm h}(t - {\bf\Omega}\cdot{\bf x}_{k}) - {\rm h}(t - {\bf\Omega}\cdot{\bf x}_{k+1})
\right]
\left(1+{\bf\Omega}\cdot{\bf n}_{k,k+1}\right).
\end{eqnarray}
In the sum over one loop successive terms in square brackets of (\ref{eq:st:nk}) cancel;
terms in (\ref{eq:st:nk}) proportional to ${\bf\Omega}\cdot{\bf n}_{k,k+1}$ cancel
with the similar terms coming from the $(N-k)$-th link for the journey in the opposite direction.

\section{Summary}

The paper investigated the response of gravitational wave detectors to different polarization modes.
The explicit expressions for the detector responses in different static configurations were given
together with their angular and frequency characteristics. We note that the explicit expressions
for the one way response can immediately be rewritten in a form which accounts for the orbital
and rotational motion of the interferometers around the Sun and they enable to construct various
time-delay combinations which are valid for any frequency in the detector's band. The time dependent
position  ${\bf x}(t)$ and orientation ${\bf n}(t)$ of the detector's arm that enter to the response
would give rise to the amplitude and phase modulation. Following e.g.\cite{Krolak04} one can
then construct the optimal filters which differ in the amplitude modulation functions (cf. Eqs.(\ref{eq:uv:sl})-(\ref{eq:uv:tc})) and study the separate detectability of the polarized signals
of the quasi-continuous sources via matched filtering. In the case of the stochastic signals the
responses define the overlap reduction function and allow to go beyond the LW approximation which
may prove necessary for the proper detection and estimation of the polarization modes of the stochastic gravitational-wave background (e.g. in cases when the lower band is restricted by the foreground of compact binaries\cite{Nishizawa10}). We also presented the frequency characteristics of the angular power of
the lowest multipole moments of the antenna pattern function for the stochastic background in the
triangular configuration of the detector. The correlation analysis shows that for the low frequencies the angular power resides in the lowest even multipole moments, $l=0,2,4$, for the self-correlated signals and in $l=4$ moment for the cross-correlated signals. Interestingly the angular power for both longitudinal modes decrease slowly with the frequency showing better directional sensitivity of the interferometer for those modes in the high-frequency regime which may influence the detectability of the putative anisotropy in the stochastic gravitational-wave background.


\section{Acknowledgments}
The work was supported in part by MNiSW Grant no. N N203 387237.

\appendix

\section{Response of the Delay Line interferometer}
\label{app:DL}
We derive here the frequency response for the plane monochromatic wave moving in the
direction ${\bf\Omega}$ for the Delay Line interferometer in the arrangement shown
in Fig.\ref{f:dlink}. To this end we consider the frequency fluctuation in the case
when the laser beam is emitted from the point ${\bf x}_a$ at the time $t_a$, reaches
the point ${\bf x}_b$ at the time $t_b$ and is detected at the point ${\bf x}_c=0$ at the
time $t$. Then the response $y_{a,b,c}(t)$ is given by
\begin{eqnarray}
\label{eq:comp}
y_{a,b,c}(t) & = & y_{b,c}(t) + y_{a,b}(t-L_{b,c})\\
&=&
-i\left[ \omega L_{b,c}{\cal T}(\omega;{\bf x}_{b,c},{\bf\Omega})F^{\pi}({\bf n}_{b,c}) +
\omega L_{a,b}{\cal T}(\omega;{\bf x}_{a,b},{\bf\Omega})F^{\pi}({\bf n}_{a,b})\,e^{-i\Delta\Phi}
\right]\,e^{i\omega t}\nonumber,
\end{eqnarray}
where the phase shift $\Delta\Phi:=\omega L_{b,c}(1 - {\bf \Omega}\cdot{\bf n}_{b,c})$
accounts for the time, $L_{b,c}$, and space, $-{\bf \Omega}\cdot{\bf x}_{b,c}$,
shifts for the journey from ${\bf x}_a$ to ${\bf x}_b$.

Using the formula (\ref{eq:comp}) for the composition of two susequent responses we can write the
response corresponding to the single round trip from the front mirror to the end mirror and back as
$-2\,i\,\omega L\,{\cal T}_{1}$ where the normalized transfer function ${\cal T}_1$ is given by
\begin{eqnarray}
\label{eq:single}
{\cal T}_{1}(\omega;{\bf x},{\bf\Omega}) & := & \frac12\left[{\cal T}(\omega;-{\bf x},{\bf\Omega}) +
{\cal T}(\omega;{\bf x},{\bf\Omega}),e^{-i \omega L (1+c)}\right]\\
& = &
\frac12e^{-i\; \omega L}\left\{
\text{sinc}[\frac{\omega L}{2}(1 + c)]e^{-\frac{i \;\omega L}{2}(-1 + c)} +
\text{sinc}[\frac{\omega L}{2}(1 - c)]e^{-\frac{i \;\omega L}{2}(1 + c)}\right\}\nonumber;
\end{eqnarray}
here $\frac12$ is the normalization factor and $c\equiv{\bf \Omega}\cdot{\bf x}/L$.
The normalized transfer function for the multiple round trip reads
\begin{eqnarray}
\label{eq:TDL}
{\cal T}_{N}(\omega;{\bf x},{\bf\Omega}) & := &
{\cal T}_{1}(\omega;{\bf x},{\bf\Omega})\;
\frac{\sin{N \omega L}}{N\sin{\omega L}}
e^{-i \omega L(N -1)}.
\end{eqnarray}
Finally we consider a Fabry-Perot cavity in the reflection mode with the reflection coefficient $\rho$ of the
front mirror inside the cavity. The normalized transfer function ${\cal T}_{FP}$ reads
\begin{eqnarray}
\label{eq:TFP}
{\cal T}_{FP}(\omega;{\bf x},{\bf\Omega}) & := &
{\cal T}_{1}(\omega;{\bf x},{\bf\Omega})\;\frac{1-\rho}{1-\rho e^{-2i \omega L}}.
\end{eqnarray}
Eqs. (\ref{eq:single}), Eqs. (\ref{eq:TDL}) and (\ref{eq:TFP}) agree with the result derived by Schilling
\cite{Schilling97} for the case of the tensorial polarization for a single arm interferometer.

Using repeatedly the formula (\ref{eq:comp}) and assuming that the measurement
of the frequency fluctuation is performed at the beam splitter at the origin
of the reference frame  the response $y_{N}(t)$ defined in ($\ref{eq:dlink}$)
can be written as:
\begin{eqnarray}
\label{eq:dlink:DLapp}
y_{N}(t)&=&-i\left[
\omega L_{bs,f}{\cal T}(\omega;{\bf x}_{f,bs},{\bf\Omega}) + \right.\\
&&
\left.
\omega L\,{\cal T}(\omega;{\bf x}_{e,f},{\bf\Omega})e^{-i \omega L_{bs,f}(1+c)}
\frac{1-e^{-2i N \omega L}}{1-e^{-2i \omega L}} +  \nonumber\right.\\
&&
\left.
\omega L\,{\cal T}(\omega;{\bf x}_{f,e},{\bf\Omega})
e^{-i\omega (L + L_{bs,f})(1+c)}
\frac{1-e^{-2i N \omega L}}{1-e^{-2i \omega L}} + \nonumber\right.\\
&&
\left.
\omega L_{f,bs}\,{\cal T}(\omega;{\bf x}_{bs,f},{\bf\Omega})
e^{-i \omega L_{bs,f}(1+c)}e^{-2i N \omega L}
\right]F^{\pi}({\bf n})e^{i\omega t}  \nonumber\\
&&
-i\,
\omega L_{em,bs}\,{\cal T}(\omega;{\bf x}_{em,bs},{\bf\Omega})
e^{-2i \omega L_{bs,f}}e^{-2i N \omega L}
F^{\pi}({\bf n}_{em,bs})e^{i\omega t}.\nonumber
\end{eqnarray}
The terms multiplying the transfer functions in the formula (\ref{eq:dlink:DLapp}) arise from
the corresponding phase shifts for the basic responses (\ref{eq:yab}) taken at different
times and space points. For a monochromatic plane wave detected at the point ${\bf x}=0$
the definition (\ref{eq:fresp}) gives $y(t)=\omega\,H(\omega)e^{i\omega t}$, thus using
the formula (\ref{eq:dlink:DLapp}) together with Eqs. (\ref{eq:single})-(\ref{eq:TFP}) we get
the frequency responses (\ref{eq:dlink:DL}), (\ref{eq:dlink:FP}) for DL and FP single-arm detectors.

\section{Angular pattern functions in the LW limit}
\label{app:uv}
In this appendix we give explicit forms of the functions $u$ and $v$ that define
angular antenna pattern functions (\ref{eq:uv}) in the LW limit for each polarization mode $\pi$.
We first express the source's basis as
\begin{eqnarray}
{\bf e}_{x'} & = &
{\bf f}_{x}\;\cos{\psi}
+ {\bf f}_{y} \sin{\psi} \\
{\bf e}_{y'} & = &
-{\bf f}_{x}\;\sin{\psi}
+{\bf f}_{y}\;\cos{\psi}\nonumber\\
{\bf e}_{z'} & = & {\bf f}_{z}
\end{eqnarray}
where the three unit vectors
$\{{\bf f}_{x},{\bf f}_{y},{\bf f}_{z}\equiv{\bf \Omega}\}$
are defined by
\begin{eqnarray}
{\bf f}_{x} & = &
{\bf e}_{x}\;\sin{\phi}
-{\bf e}_{y} \cos{\phi} \\
{\bf f}_{y} & = &
-{\bf e}_{x}\;\cos{\phi}\cos{\theta}
-{\bf e}_{y}\;\sin{\phi}\cos{\theta}
+{\bf e}_{z}\;\sin{\theta} \nonumber\\
{\bf f}_{z} & = &
-{\bf e}_{x}\;\cos{\phi}\sin{\theta}
-{\bf e}_{y}\;\sin{\phi}\sin{\theta}
-{\bf e}_{z}\;\cos{\theta}\nonumber.
\end{eqnarray}
For the arm oriented along ${\bf n}_{a,b}$  the functions $u$ and $v$ then read
\begin{eqnarray}
\label{eq:uv:sl}
\text{ scalar longitudinal:}
&&
u_{sl} = 1/2 ({\bf n}_{a,b}\cdot{\bf f}_{z})^2\\
\text{ scalar transversal:}
\label{eq:uv:st}
&&
u_{st} = 1/2\left[ ({\bf n}_{a,b}\cdot{\bf f}_{x})^2 +
({\bf n}_{a,b}\cdot{\bf f}_{y})^2\right]\\
\text{ vectorial $x$:}
&&
u_{vx} = ({\bf n}_{a,b}\cdot{\bf f}_{x})({\bf n}_{a,b}\cdot{\bf f}_{z})\\
&&
v_{vx} = ({\bf n}_{a,b}\cdot{\bf f}_{y})({\bf n}_{a,b}\cdot{\bf f}_{z})\nonumber\\
\text{ vectorial $y$:}
&&
u_{vy} = ({\bf n}_{a,b}\cdot{\bf f}_{y})({\bf n}_{a,b}\cdot{\bf f}_{z})\\
&&
v_{vy} = -({\bf n}_{a,b}\cdot{\bf f}_{x})({\bf n}_{a,b}\cdot{\bf f}_{z})\nonumber\\
\text{ tensorial $+$:}
\label{eq:uv:tp}
&&
u_{tp} = 1/2\left[ ({\bf n}_{a,b}\cdot{\bf f}_{x})^2 -
({\bf n}_{a,b}\cdot{\bf f}_{y})^2\right]\\
&&
v_{tp} = ({\bf n}_{a,b}\cdot{\bf f}_{x})({\bf n}_{a,b}\cdot{\bf f}_{y})
\nonumber\\
\text{ tensorial $\times$:}
\label{eq:uv:tc}
&&
u_{tc} = ({\bf n}_{a,b}\cdot{\bf f}_{x})({\bf n}_{a,b}\cdot{\bf f}_{y})\\
&&
v_{tc} = 1/2\left[ ({\bf n}_{a,b}\cdot{\bf f}_{y})^2 -
({\bf n}_{a,b}\cdot{\bf f}_{x})^2\right]
\nonumber
\end{eqnarray}
%
\section{M, DLM and FPM interferometers}
\label{ch:DLM}
In this appendix we present the responses for the Michelson Delay Line and
Michelson Fabry-Perot interferometers. We assume that the emitter, beam splitter,
front and end mirrors of the first arm are aligned along a unit vector ${\bf n}_1$;
the second arm lies along a unit vector ${\bf n}_2$. The response for the DLM detector then reads
\begin{eqnarray}
\label{eq:DLM:full}
y_{DLM}(t) & = &
\left\{
-\frac{e^{-i \omega\left[ L_1 (2 N + c_1) + (1 + c_1) L_{bs,f1}\right]}}
{\left(1 - c_1^2\right) \left( 1 - e^{2 i \omega L_1}\right)}
\left[(1 - c_1) e^{i\omega (2 + c_1) L_1}
+2 c_1 e^{i \omega L_1} - \left( 1 + c_1 \right) e^{i \omega L_1 c_1}\right]\times\right.\nonumber\\
&&
\left(1 - e^{2 i N \omega L_1}\right)
+
\frac{1}{{1 - c_1}}
\left( e^{-2 i \omega\left(N L_1 + L_{bs,f1}\right)}
-e^{-i \omega\left[(1 + c_1) L_{bs,f1} + 2 N L_1\right]} \right) +\nonumber\\
&&
\left.
\frac{1}{1 + c_1}
\left(
e^{-i \omega(1 + c_1) L_{bs,f1}} - 1
\right)
\right\}F^{\pi}({\bf n}_1)\;e^{i\omega t}
-
\qquad \left( 1\leftrightarrow 2 \right) \nonumber\\
&&
+\frac{\left(1 - e^{- i \omega(1 - c_1) L_{e,bs}}\right)}{1 - c_1}
e^{-2 i\omega \left[N \left( L_1 + L_2\right) + L_{bs,f1} + L_{bs,f2}\right]}\nonumber\\
&&
\left(e^{2 i \omega\left(N L_1 + L_{bs,f1}\right)} -
e^{2 i \omega \left(N L_2 + L_{bs,f2}\right)}
\right)F^{\pi}({\bf n}_1)\;e^{i\omega t}
\end{eqnarray}

For the FPM detector we have
\begin{eqnarray}
\label{eq:FPM:full}
y_{FPM}(t) & = &
\left\{
-\frac{(1 + \rho)}{\left(1 - c_1^2\right) \left(\rho - e^{2 i \omega L_1}\right)}
\left[c_1
\left(e^{i\omega (1 + c_1) L_1}+e^{i \omega(3 + c_1) L_1}-2 e^{2 i \omega L_1}\right)+\right.\right.\\
&&
\left.
e^{i \omega(1 + c_1) L_1} - e^{i \omega(3 + c_1) L_1}
\right]
e^{-i \omega(1 + c_1) \left(L_1 + L_{bs,f1}\right)}
+ \frac{-1 + e^{-i \omega(1 + c_1) L_{bs,f1}}}{1 + c_1}-\nonumber\\
&&
\left.
\frac{e^{-2 i \omega L_{bs,f1}}
\left(1 - e^{i \omega(1 - c_1) L_{bs,f1}}\right)
\left(1 - e^{2 i \omega L_1} \rho \right)}{(1 - c_1) \left(\rho - e^{2 i \omega L_1}\right)}
\right\}F^{\pi}({\bf n}_1)\;e^{i\omega t}
\quad - \quad \left( 1\leftrightarrow 2 \right)
\nonumber\\
&&
+ \frac{\left(1 - e^{i (c_1-1) x_{em,bs}}\right)}
{1 - c_1}
\left(\frac{e^{-2 i x_{bs,f1}}
\left(1 - e^{2 i x_1} \rho \right)}{\rho - e^{2 i x_1}} -
\frac{e^{-2 i x_{bs,f2}} \left(1 - e^{2 i x_2} \rho \right)}
{\rho - e^{2 i x_2}}\right)\nonumber\\
&&
F^{\pi}({\bf n}_1)\;e^{i\omega t}\nonumber
\end{eqnarray}
In Eqs. (\ref{eq:DLM:full}) and (\ref{eq:FPM:full}) $L_1$ and $L_2$
denote the lengths between two mirrors in arm1 and arm2 respectively,
$c_1\equiv{\bf \Omega}\cdot{\bf n}_1$, $c_2\equiv{\bf \Omega}\cdot{\bf n}_2$.



\end{document}